# Local sequence-structure relationships in proteins

Tatjana Škrbić[1,2,*], Amos Maritan[3], Achille Giacometti[2] and Jayanth R. Banavar[1]

1. Department of Physics and Institute for Fundamental Science, University of Oregon, Eugene, OR 97403, USA

2. Dipartimento di Scienze Molecolari e Nanosistemi, Università Ca' Foscari Venezia, Campus Scientifico, Edificio Alfa, via Torino 155, 30170 Venezia Mestre, Italy

3. Dipartimento di Fisica e Astronomia, Università di Padova and INFN, via Marzolo 8, 35131 Padova, Italy

**Abstract**

**We seek to understand the interplay between amino acid sequence and local structure in proteins. Are some amino acids unique in their ability to fit harmoniously into certain local structures? What is the role of sequence in sculpting the putative native state folds from myriad possible conformations? In order to address these questions, we represent the local structure of each $C_\alpha$ atom of a protein by just two angles, $\theta$ and $\mu$, and we analyze a set of more than 4000 protein structures from the PDB. We use a hierarchical clustering scheme to divide the 20 amino acids into six distinct groups based on their similarity to each other in fitting local structural space. We present the results of a detailed analysis of patterns of amino acid specificity in adopting local structural conformations and show that the sequence-structure correlation is not very strong compared to a random assignment of sequence to**

**structure. Yet, our analysis may be useful to determine an effective scoring rubric for quantifying the match of an amino acid to its putative local structure.**

**Significance statement:** We present a quantitative study of the emergent constraints of sterics, the chain topology, and the quantum chemistry on local protein native state structures measured in a simple representation. We present two main classes of results: the propensity of amino acids to occupy certain local structures and a grouping of amino acids based on their similarity in hosting local structures.

It is known that there are just a few important principles (1-6) that drive the folding process of a protein: the requirement of avoiding steric overlaps in both the folded and unfolded states, the lower conformational entropy in the folded state than in the unfolded state, the hydrophobic effect favoring a compact conformation that is able to expel water from the core of the folded state and the delicate balance of hydrogen bonds with the solvent and within the protein backbone that can tip the energetic balance between the unfolded and folded state. The fundamental issue is how nature has effectively explored the astronomically large sequence space through evolution to make proteins the molecular target of natural selection.

Here we characterize the native state folds within a simple coarse-grained representation and elucidate the role, if any, played by the repertoire of amino acids in fitting into one of these local geometries. We model a chain by just its $C_\alpha$ atoms and follow the coordinate representation shown in Figure 1. With the knowledge of the preceding $C_\alpha$ locations, we specify the position of a given $C_\alpha$ atom by three coordinates (7), the bond length, b, and two angles, $\theta$ and $\mu$. $\theta$ is the bending angle at the given $C_\alpha$ location, whereas $\mu$ is the angle between successive binormals (Figure 1). The binormal associated with a specific consecutive triplet of $C_\alpha$ atoms is the unit vector perpendicular to the plane of the triplets. The tangent, the normal, and the binormal, all at the middle $C_\alpha$ atom, form a right-handed Cartesian coordinate system. This coordinate system was introduced by Rubin and Richardson in a paper describing the Byron bender that allowed for a simple construction of protein $C_\alpha$ models (8,9).

Our analysis is carried out with a set of more than 4000 experimentally determined protein native state structures. Starting from the Top 8000 set proteins of the Richardson laboratory (10,11) with 70% homology level, we excluded all structures with missing atoms in the protein backbone, yielding a set of 4416 protein native state structures that we used for our analysis (the same set was employed in Ref. 7) (see Table S1 in Supplementary Information). We successfully validated our analysis using 478 proteins from the Dunbrack data set (12), this time with a maximum sequence homology level of 20%. There were 205 proteins in common between the Richardson and Dunbrack sets that we employed. We carried out the ($\theta$, $\mu$) analysis for both the Richardson and Dunbrack data sets and obtained virtually identical results with the Dunbrack data being understandably more sparse. We present here the detailed analysis for just the much larger Richardson data set.

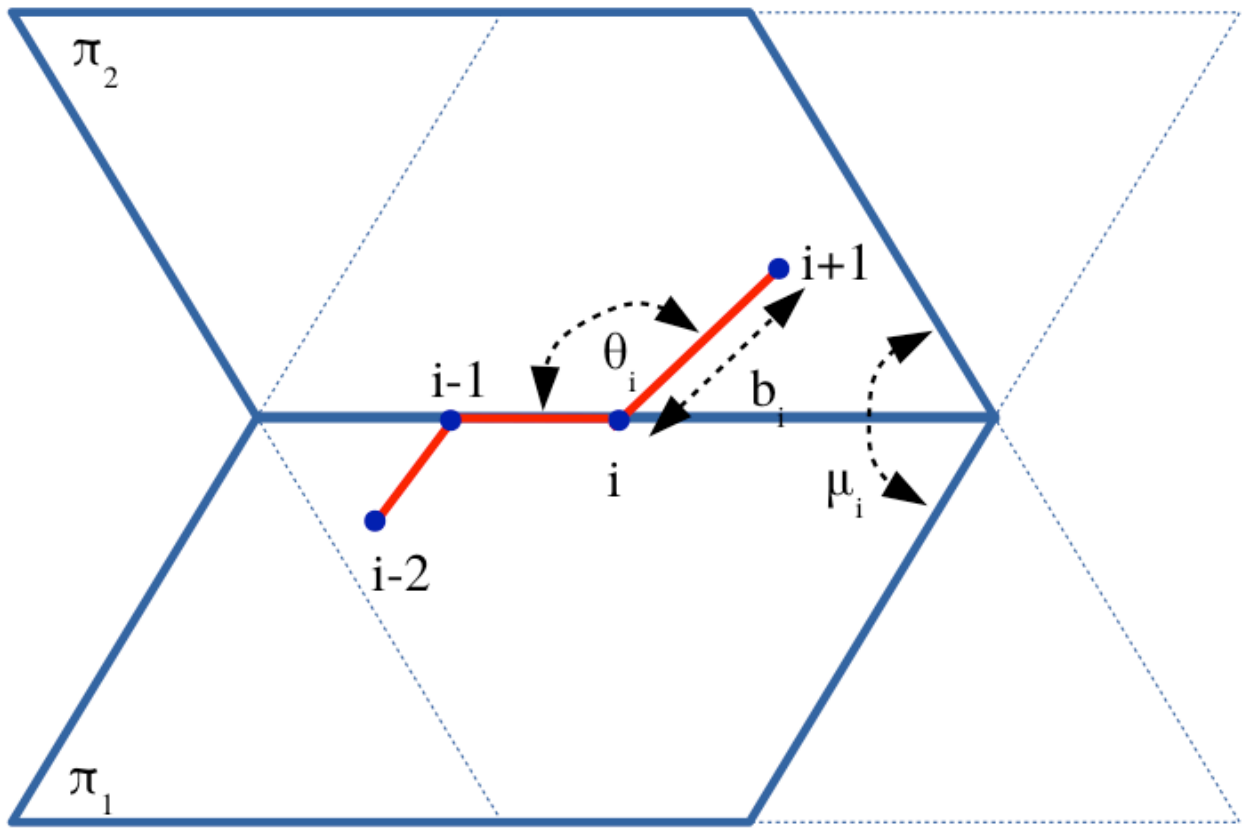

**Figure 1: Definition of coordinate system.** The bond length b at location i, $b_i$, is the distance between the points i and (i+1). The angle $\theta_i$ is the angle subtended at i by points (i-1) and (i+1) along the chain. The third coordinate $\mu_i$ is the dihedral angle between the planes $\pi_1$ and $\pi_2$ formed by [(i-2), (i-1), i] and [(i-1),i,(i+1)] respectively and is the angle between the binormals at (i-1) and i. Knowledge of the coordinates of the previous three points (i-2,i-1,i) and the three variables ($b_i$, $\theta_i$, $\mu_i$) are sufficient to uniquely specify the coordinates of the point (i+1).

A simplification arises because the vast majority of bond lengths is nearly constant (Figure 2). Figures 2a and a blown up version, Figure 2b, depict histograms of bond lengths with two peaks centered around 3.81Å and 2.95Å. The shorter bonds are associated with a Ramachandran angle ω (1) around 0 degrees (13) (Figure 2c). Because the fraction of short bonds is relatively small (0.3%), our analysis here is carried out with <u>all</u> $C_\alpha$ positions, each characterized by a bond length, the θ and μ angles and the amino acid identity. An analysis of the amino acids associated with just the short bonds shows the preponderance of glycine in the first position and proline in the second position (because of the low barrier for transitioning between its cis and trans conformations).

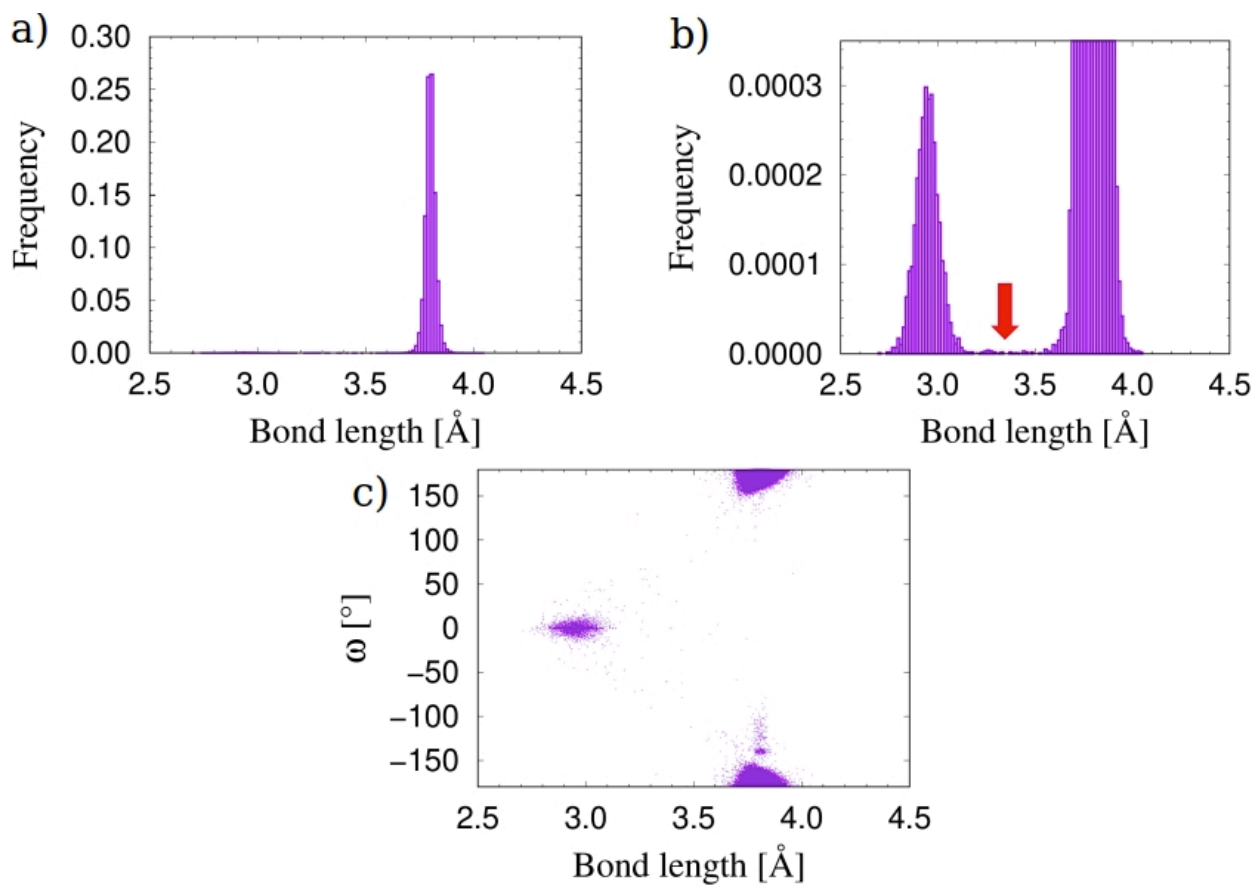

**Figure 2: Distribution of bond lengths.** Figure 2a shows a histogram of bond lengths in our data set. A blown up version in Fig 2b shows that the distribution is bimodal with short bonds (centered around 2.95Å) and long bonds (centered around 3.81Å). The red arrow is the length

we use for partitioning the bonds into the short and long categories. Figure 2c shows the link between the Ramachandran ω angle (1,13) and the bond length.

For a non-interacting phantom chain, one obtains a uniform distribution of points in the $(\theta, \mu)$ plane (not shown as a figure). As a benchmark, we studied, using Wang-Landau Monte Carlo simulations (14), a simple self-avoiding polymer chain model comprised of 40 unit diameter tangent spheres (tethered hard spheres) subject to a self-attraction between sphere centers located within a distance of 2 units of each other. Figure 3a and 3b show a cross plot in the $(\theta, \mu)$ plane of 17 conformations in the coil phase adopted by the chain at high temperatures and for 17 low energy conformations, respectively. The situation is dramatically different for proteins compared to a standard self-avoiding polymer model. Figure 3c is the $(\theta, \mu)$ cross plot for the protein data set with a highly selective occupancy of $(\theta, \mu)$ space (a version of this graph was presented earlier in Ref. 7).

We binned the data in Figure 3c into squares of width 5° along θ (24 bins in the range 60°-180°) and 5° along μ (72 bins spanning the range from 0° to 360°) to determine the three highest density regions. These density peaks are shown in the figure as black X's along with three larger squares of size 10°×10° around them. They are identified as helices (blue region with black X at θ = 92.5° and μ = 47.5°), β-strands (red region with black X at θ = 122.5° and μ = 192.5°), and loops (green region with black X at θ = 92.5° and μ = 242.5°) with 184382, 16372, and 10974 points respectively. The density of points in the α-helix peak is approximately 20 times that of loops and β-strands but the loop and β-strand regions are more spread out than the helical region. The other populated regions in the (θ, μ) plane correspond to variants of helices and β-strands and the loops that link them together in the native state structure.

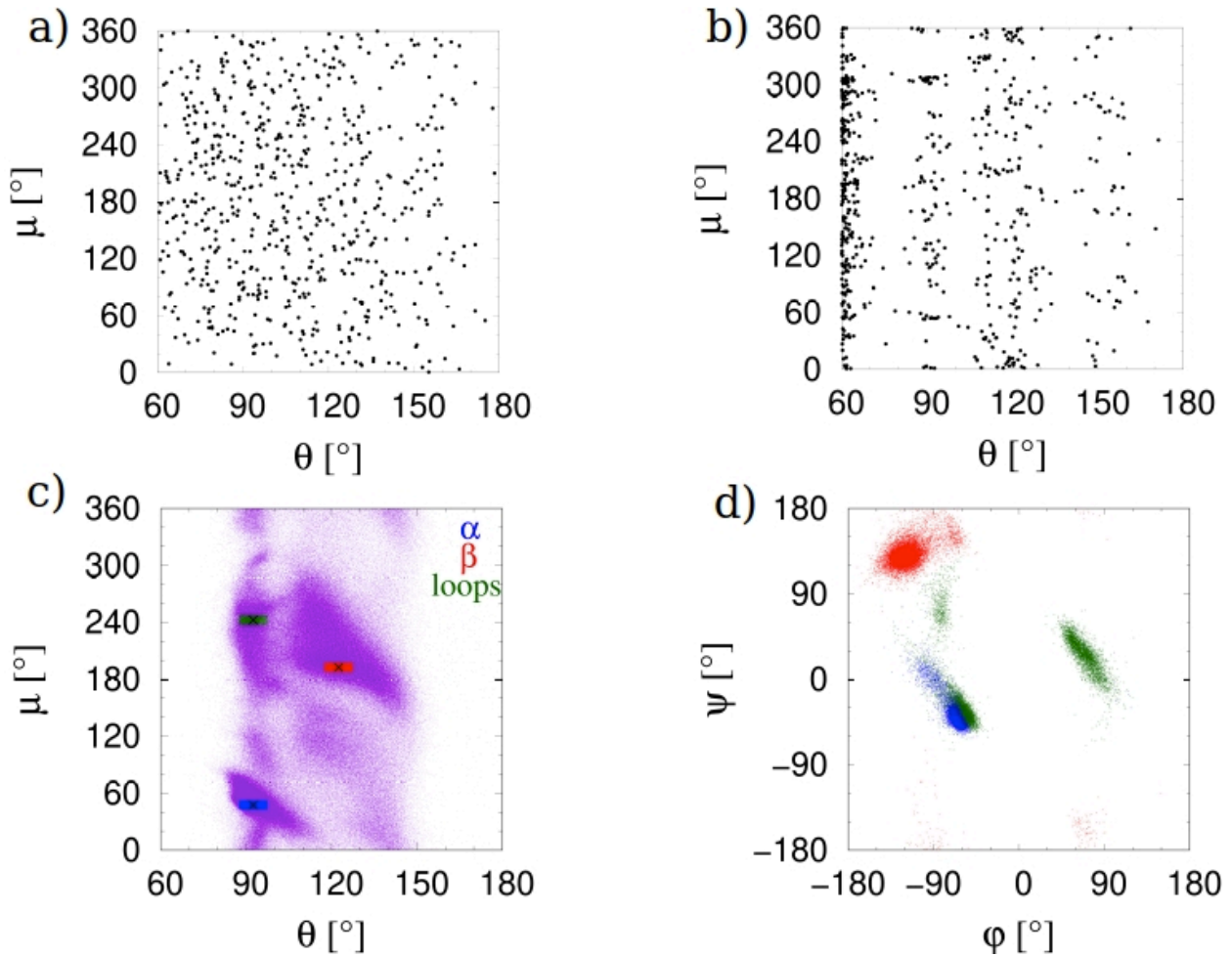

**Figure 3: Local structure representation.** a) (θ, μ) cross plot for the high temperature coil phase of tethered hard spheres. The only constraint here is the requirement of self-avoidance of the spheres. The points are scattered across the plane with no θ angle less than 60° (a steric constraint) and few almost straight line triplets with a θ near 180°. b) (θ, μ) cross plot for low energy states of tethered hard spheres. Here again one observes no θ angles below 60° and favored θ angles of 60°, 90°, 120°, and 150° showing that the order favors a face-centered-cubic packing locally, which would be appropriate for the close packing of untethered spheres. c) (θ, μ)

plot for the Richardson data set comprising 4416 proteins and 972519 residues (purple points). The three highlighted regions correspond to density peaks related to $\alpha$-helices (blue region $\theta$ = 92.5° and $\mu$ = 47.5°), β-strands (red region $\theta$ = 122.5° and $\mu$ = 192.5°), and loops (green region $\theta$ = 92.5° and $\mu$ = 242.5°)  (d) Plot of the Ramachandran ($\varphi$, $\psi$) angles for the highlighted regions in Figure (c).

It is important to note that the angles $\theta$ and $\mu$ are distinct from the Ramachandran (1) angles, which require the knowledge of the locations of backbone atoms besides those of the $C_\alpha$ atoms. The ($\theta$, $\mu$) pair is a coarse grained representation of the Ramachandran angles and can be useful to describe a generic chain conformation and employed in models of statistical mechanics (15). In fact, knowing a sequence of Ramachandran angles, one can derive the values of $\theta$ and $\mu$. The inverse process of determining the Ramachandran angles from the ($\theta$, $\mu$) values does not have a unique solution. For the $C_\alpha$ atoms in the interior of all 4416 proteins, we measured the ($\theta$, $\mu$) as well as the Ramachandran ($\varphi$, $\psi$) angles. We illustrate the relationship between the two coordinate systems in Figure 3d. We plot the three colored regions (blue, red and green) of dense points in Figure 3c, but this time expressed as the ($\varphi$, $\psi$) Ramachandran angles color coded in the same manner as in the ($\theta$, $\mu$) plot.  Note that the closely packed points in the ($\theta$, $\mu$) plot are more dispersed in the Ramachandran plot sometimes occupying non-contiguous regions. This is because $\theta$ and $\mu$ depend on more than one set of Ramachandran angles and the relationship is complicated and non-linear.

There are four important earlier papers that our work builds on. Rackovsky and Scheraga (16) considered a torsion-curvature plot (distinct from but related to the plot we studied) for 22 protein structures for two different structural groups (helices + bends and extended strands) and the amino acids present therein. Levitt (17) analyzed 13 proteins and considered a ($\theta$, $\mu$) plot similar to ours except that the definition of mu was shifted by one $C_\alpha$ position in the backward direction compared to our definition. Our own definition was motivated by defining $\theta$ and $\mu$ at a given site i that would determine the coordinates of the (i+1)-th $C_\alpha$ coordinate. Importantly, Levitt determined an approximate empirical relationship between his $\theta$ and $\mu$ to elucidate approximate potentials for folding simulations.

Oldfield and Hubbard (18) considered two successive $\theta$ angles and one $\mu$ angle (defined for a bond joining the two $C_\alpha$ atoms) for a set of 83 protein structures and carried out a comprehensive study of local conformational space (but not amino acid preferences) recognizing that the two major building blocks of protein native state structures, helices and strands, are repetitive conformations. DeWitte and Shakhnovich (19) considered 87 protein structures with a goal of deducing the pairwise potentials, in the spirit of Miyazawa and Jernigan, for the formation of secondary structures in protein simulations based on a cross-plot of two successive $\mu$ angles (this time again defined as bond variables rather than at a site) and employing Levitt's empirical relationship. Finally, the approach of Bahar, Kaplan, and Jernigan (20) is most similar to ours. They do have a ($\theta$, $\mu$) plot just like ours except that their $\mu$ definition is shifted by one position compared to ours. They employed 302 protein structures for their analysis, they carried out an amino acid propensity estimate like we do, and they successfully developed short-ranged (along the sequence) rotational potentials for single amino acids.

In essence, our work here builds on these earlier advances. The principal distinctions are the definition of μ - our μ is defined at a site not at a bond, it is shifted with respect to other definitions, and the number of protein structures we employ, many decades after the earliest work, is understandably larger and comprises over 4000 experimentally determined and curated protein structures. Our goal in this paper is not to extract effective potentials but rather analyze, more generally, sequence-local structure relationships. Furthermore, we seek to group the 20 amino acids into distinct groups in terms of their similarity to substitute for each other in local conformational space.

Figure 4 shows histograms of θ and μ values and evidence for a clear correlation between the average values of θ and the average value of μ among all proteins. Table 1 presents data on the amino acid occurrence probability and the degree of localization in (θ, μ) space. For each amino acid, we measured the inverse participation ratio (IPR) defined as

$$IPR = \frac{(\sum_{i=1}^{N} x_i^2)^2}{\sum_{i=1}^{N} x_i^4} \quad (1)$$

where $x_i$ denotes the normalized density of occupancy of the i-th bin in (θ, μ) space and the total number of bins N=1728. An IPR value of 1 indicates perfect localization in just one bin whereas the largest possible value of the IPR is N=1728 for a uniform occupancy of all 1728 bins.

A perfect localization (IPR=1) is indicative of an amino acid that is always associated with the same local structure leading to a perfect sequence-structure relationship. The most localized amino acid is LEU (IPR=2.70) while the least localized is PRO (IPR=83.28). Figure 5 shows the occupancies of the (θ, μ) space of amino acids LEU and PRO. Interestingly, even the most localized amino acid, while being largely concentrated in just a few squares, is yet spread out over many squares indicating that there is no strong selection of local structure by amino acid identity.

We carried out an analysis of triplet amino acids identities of all the 324 tight bends with θ angles less than 80°. The smallest θ angle in the data set has a value of 59.98° and the corresponding amino acid triplet is GLY-GLN-ASP. These tight turns (i-1,i,i+1) have no selectivity in μ angles. However, there is indeed a sequence-structure relationship with (GLY or SER) accounting for a total of 34% occupancy in the i-1 position, (PRO or SER) having 31% residency in site i, and (ALA or SER) accounting for 21% in site (i+1).

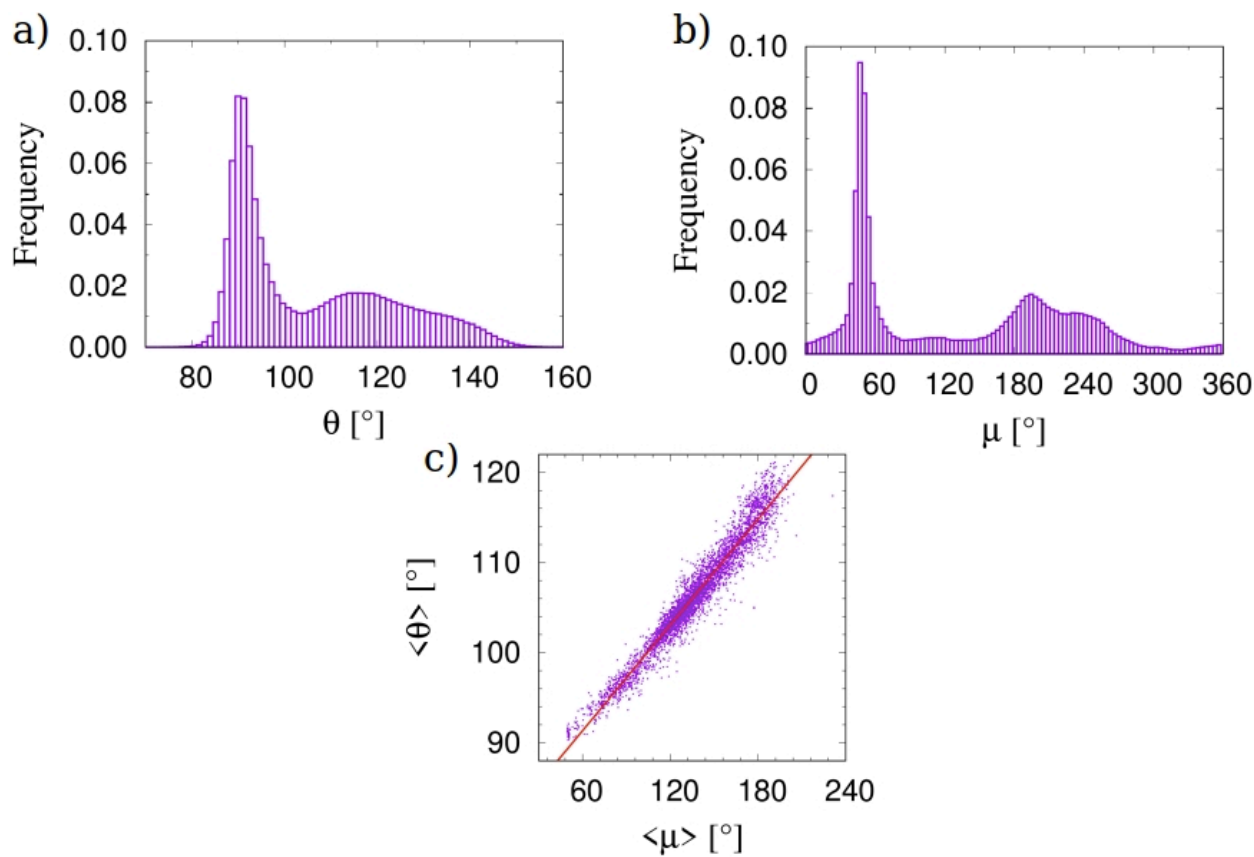

**Figure 4: a) and b). Histograms of θ and μ values showing a multi-peaked structure. c). A plot of the average value of θ versus the average value of μ for all 4416 proteins showing a tight correlation with a Pearson correlation coefficient of 0.97.** This may be readily understood by noting that a protein structure is primarily composed of helices and sheets with varying fractions depending on the protein being considered. The θ-μ values for an α-helix are both smaller than those of a β-strand leading to the correlation. Note that the standard deviations (not shown) are large because of the relatively large width in angle space of the regions.

| Amino acid type | Fraction [%] | Inverse Participation Ratio (IPR) |
|---|---|---|
| ALA | 8.53 | 3.28 |
| ARG | 4.84 | 3.24 |
| ASN | 4.42 | 4.53 |
| ASP | 5.96 | 4.60 |
| CYS | 1.36 | 3.69 |
| GLU | 6.48 | 3.25 |
| GLN | 3.61 | 3.30 |
| GLY | 7.90 | 11.61 |
| HIS | 2.32 | 4.31 |
| ILE | 5.62 | 2.93 |
| LEU | 8.79 | 2.70 |
| LYS | 5.70 | 3.43 |
| MET | 2.02 | 2.95 |
| PHE | 4.04 | 4.06 |
| PRO | 4.59 | 83.28 |
| SER | 5.88 | 5.14 |
| THR | 5.58 | 4.75 |
| TRP | 1.52 | 3.99 |
| TYR | 3.61 | 4.25 |
| VAL | 7.23 | 3.77 |

**Table 1: Frequency of 20 amino acids in the set of 4416 proteins (second column) and a measure of the localization of each amino acid in ($\theta$, $\mu$) space (third column).**

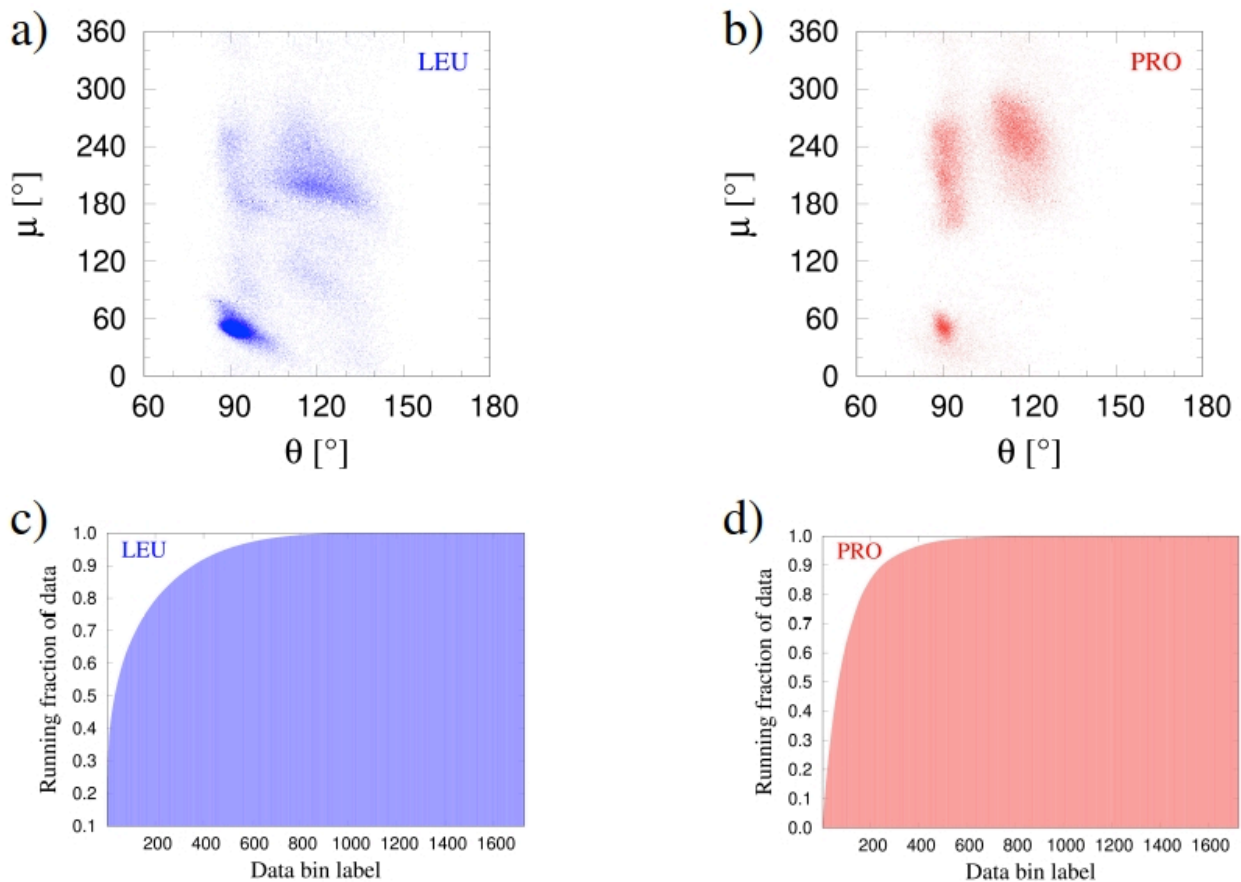

**Figure 5: Occupancy pattern of amino acids LEU and PRO in (θ, μ) space.** a) and b) depict the locations of the two amino acids. LEU is the most localized amino acid (IPR=2.70) whereas PRO has the largest IPR=83.28 value among the amino acids and is spread out the most. A rank ordered normalized occupancy fraction of the two amino acids is shown in c), and d). The number of bins needed to account for 50% and 90% occupancy for the two amino acids are LEU – 33 and 356, and PRO – 66 and 248, respectively.

We studied histograms of the θ and μ values associated with each of the twenty amino acids. The distributions are roughly equally wide and substantially independent of amino acid identity (see Figures 6 and 7).

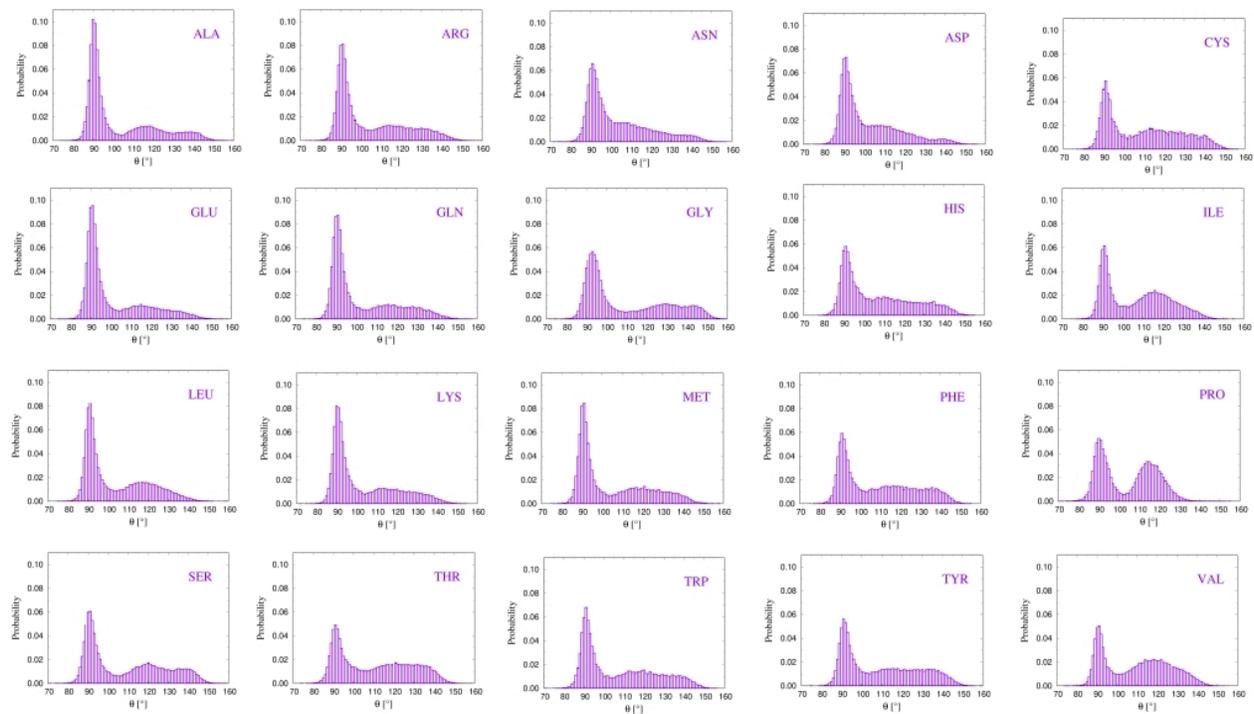

**Figure 6: Histograms of the θ values for each of the twenty amino acids.** While the shapes of the histograms vary from amino acid to amino acid, the ranges are mostly independent of amino acid identity. PRO is a bit of an outlier with a somewhat lower upper cut-off value of θ.

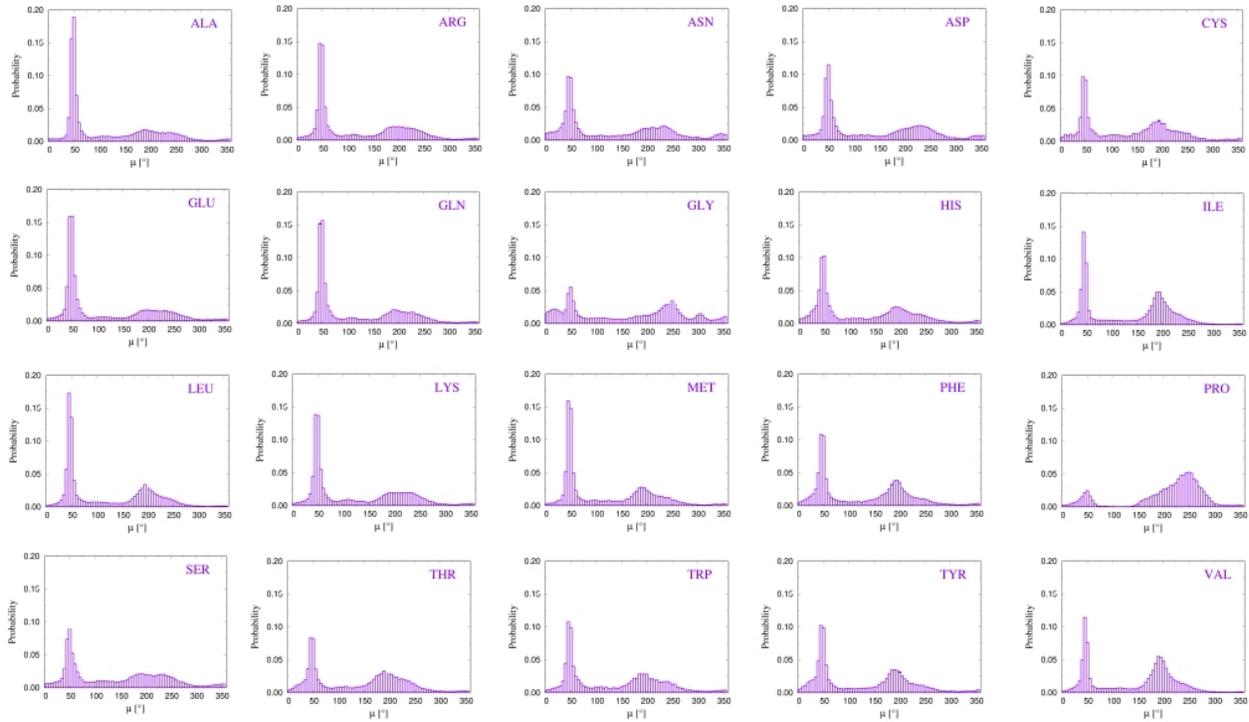

**Figure 7: Histograms of the μ values for each of the twenty amino acids.** Even though the shapes of the histograms vary from amino acid to amino acid, the ranges are mostly independent of amino acid identity.

Unlike the α-helix region associated with tight local packing and hence a relatively small variation in the θ angle, there is a range of θ values associated with the β-strand region. We carried out sequence analyses of the β-strands to understand whether there is an amino acid selection principle for θ. We selected the (θ, μ) subspace consisting of μ values in the range from 175° to 185° (±5° degree interval around the ideal value of 180°) and of θ angles in the range from 105° to 145°. We divided up the relevant range of θ angles into 40 bins of width 1°. Again, we measure the IPR defined in Eq.(1) with N=40 in this case. The extreme values of the IPR are 16.08 for the most localized amino acid, PRO, and 31.46 for the most spread out amino acid, ASP (see Figure 8). The average θ value and its standard deviation for all amino acids in the β-region is 128.0° and 9.5° respectively.

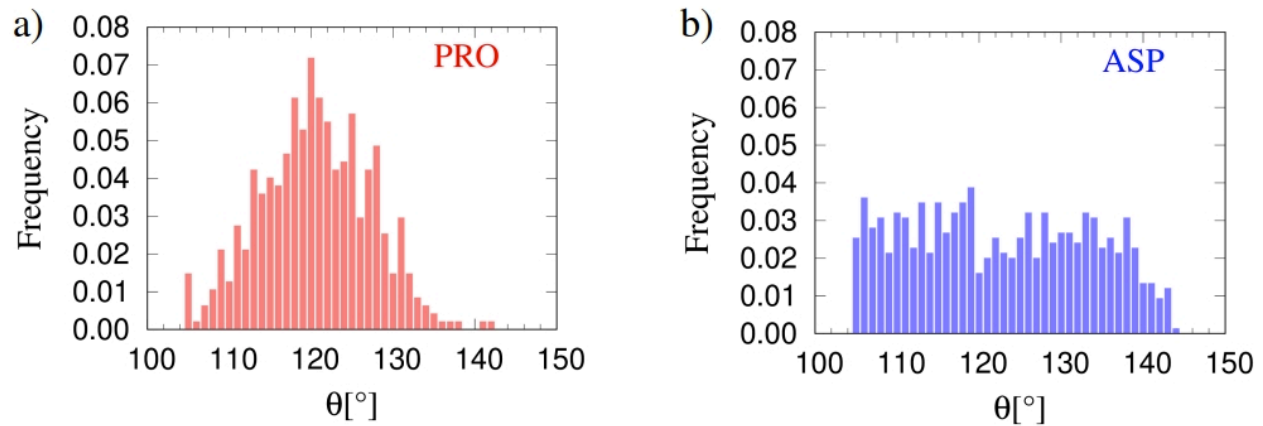

**Figure 8: Distribution of θ angles in the β- region for PRO (a) and ASP (b).** PRO is the most localized amino acid, yet exhibits some spread of θ angles.

We also studied the identities of the 210 pairs of amino acids (and their associated side chain sizes) located at sites i-1 and i+1 (these side chains stick out in roughly the same direction with a possibility of steric clashes) flanking site i in the β-region. We considered only those statistically significant pairs (i-1,i+1) which occurred at least 162 times (estimated as the total number of pairs divided by 210) with beads i-1, i, i+1 all lying in the β-strand region and divided the θ range again into 40 equally spaced bins. The number of amino acid pairs that met the 162 threshold was 52 out of the 210 pairs. We find that all pairs are spread out in θ values. The most localized pair among these was ALA-THR with an IPR of 10.51 and the most spread out pair was PHE-PRO with an IPR of 22.77 (see Figure 9 for histograms of θ values associated with these pairs). A cross plot of the mean van der Waals diameter of a pair and its average θ value (not shown) results in a weak correlation and an overall negative trend. All these results indicate that the sequence does not play a significant role in determining the θ angle associated with a β-strand.

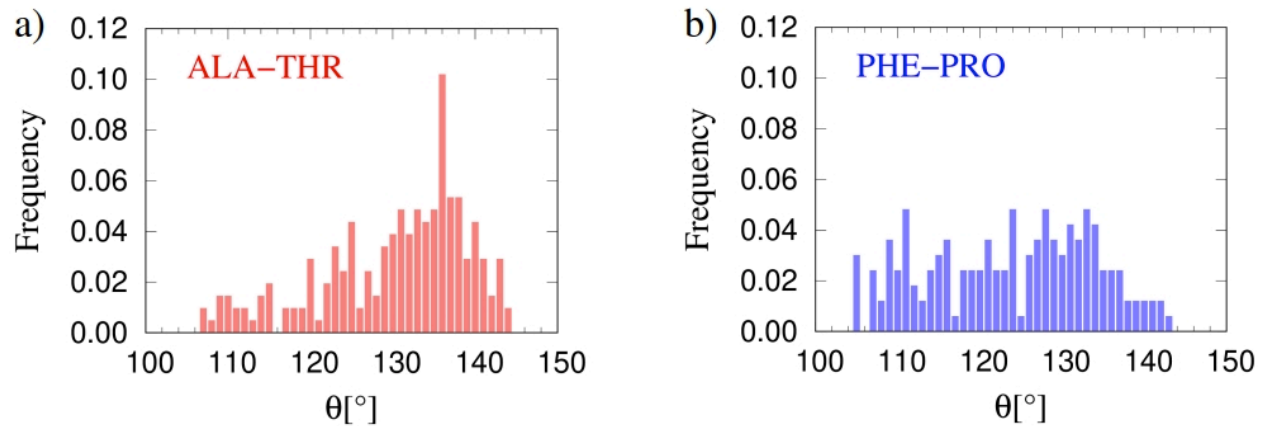

**Figure 9: Distribution of θ angles in the β-region for (ALA-THR) and (PHE-PRO) amino acid pairs in positions (i-1,i+1) respectively.** ALA-THR is the most localized pair in θ space, yet is spread out. PHE-PRO is the most spread out pair.

We carried out simple sequence analyses of the loop region as well, to understand whether there is a selection principle for the value of the μ angle. We select the (θ, μ) subspace consisting of θ angles in the range from 87.5° to 97.5° (±5° interval around the value 92.5°, identified as the peak density green region in Figure 3c) and μ values in the range from 90° to 360° to ensure that there is no overlap with the α-helix region. We divided up the range of μ angles into 54 bins of width 5°. We measured the IPR value for the 20 amino acids and we find that the most localized amino acid is GLY with a value of 8.49, whereas the most delocalized amino acid is PHE with an IPR equal to 28.42 (see Figure 10). Note that μ=180° and 360° correspond to planar configurations of 4 consecutive $C_\alpha$ atoms, with the former corresponding to zig-zagging and the latter to rotation in the same sense.

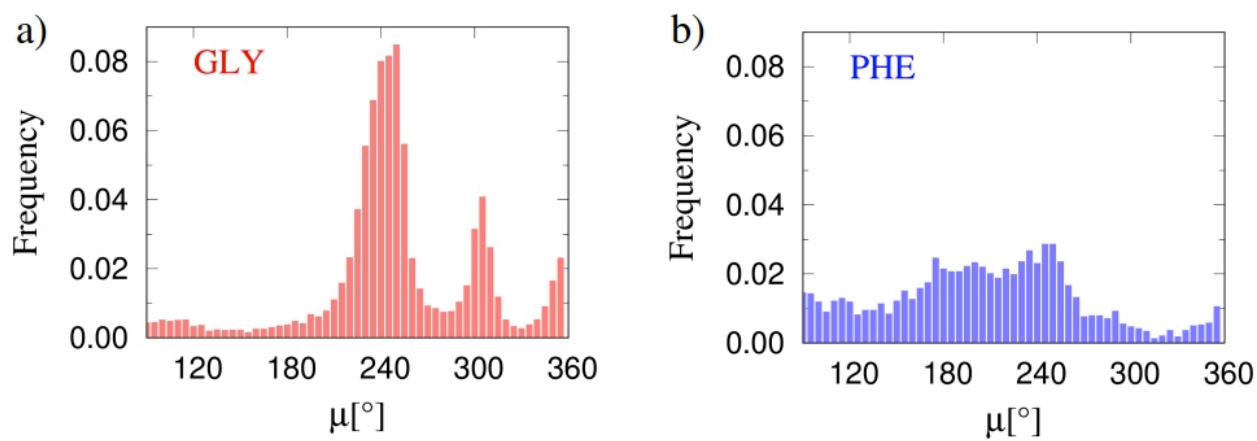

**Figure 10: Distribution of μ angles in the loop region for GLY and PHE.** GLY is the most localized amino acid, yet exhibits a spread of angles.

Based on the normalized density of occupancy of the amino acids in (θ, μ) space, one can assess the mutual similarity of the 20 amino acids by measuring the Cartesian distance between the 190 pairs of amino acids, which serves as a proxy of similarity. We have employed the Bhattacharyya coefficient (21) in order to calculate the degree of closeness of the (θ,μ) distributions of amino acids. We carried out hierarchical clustering by rank-ordering the closeness – the two closest amino acids were placed into a single group thereby now having effectively 19 groups of amino acids. This procedure was repeated recursively to reduce the effective groups of amino acids by one each time. A natural stopping point for this hierarchical clustering is when there is a relatively large jump in the measure of closeness of the remaining groups. The result of this analysis is shown in Figure 11 and yields 6 different groups comprising 7, 7, 2, 2, 1, and 1 amino acids each. Figure 12 shows the occupancy in (θ, μ) space of the six amino acid groups.

Group A: ALA - ARG - GLN - GLU - LEU – LYS - MET
Group B: CYS - HIS - PHE - SER - THR - TRP - TYR
Group C: ILE - VAL
Group D: ASN - ASP
Group E: GLY
Group F: PRO

ARG — LYS
(ARG, LYS) — GLN
(ARG, GLN, LYS) — GLU
ALA — (ARG, GLN, GLU, LYS)
(ALA, ARG, GLN, GLU, LYS) — MET
(ALA, ARG, GLN, GLU, LYS, MET) — LEU

PHE — TYR
HIS — (PHE, TYR)
(HIS, PHE, TYR) — THR
(HIS, PHE, THR, TYR) — TRP
CYS — (HIS, PHE, THR, TRP, TYR)
(CYS, HIS, PHE, THR, TRP, TYR) — SER

ILE — VAL
ASN — ASP

**Figure 11: Clustering of amino acids into groups.** The 6 amino acid groups obtained based on their similarity in occupying the local structural ($\theta$, $\mu$) space are shown. 6 is a natural choice because the closeness for the next collapse into five groups is approximately twice as large as the previous closeness measure. A 5 member group would result in the merger of the two largest groups, Group A and Group B. If one were to retain seven groups, SER would detach from Group B and remain isolated as its own group. The sequences of hierarchical clustering for the first four groups A (blue), B (red), C (purple) and D (green) is shown with the link thickness quantitatively representing the closeness measure.

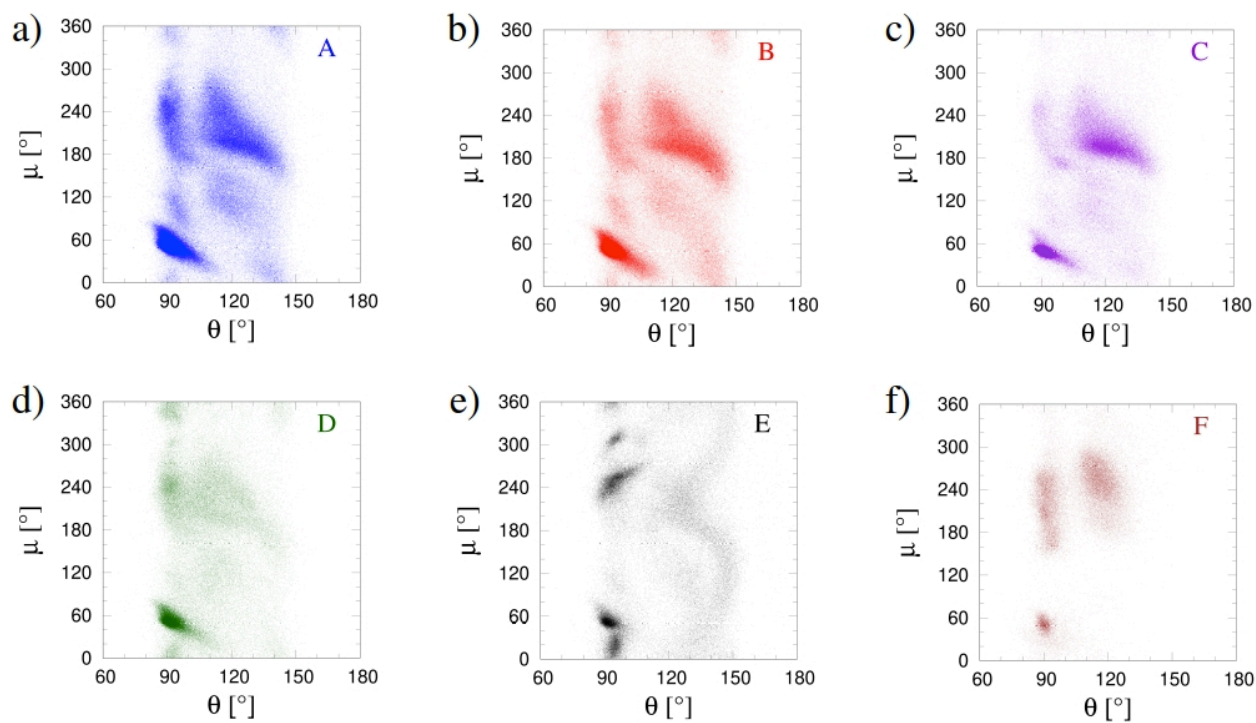

**Figure 12: Occupancy of the six amino acid groups in ($\theta$, $\mu$) space.** Groups A and B are somewhat similar with the main difference being the relative weights of the $\alpha$-helix and $\beta$-strand regions. The most distinctive groups are E and F corresponding to GLY and PRO respectively. We remind the reader (see Figure 3c) that the density peaks occur at ($\theta = 92.5°$ and $\mu = 47.5°$) for $\alpha$-helices, ($\theta = 122.5°$ and $\mu = 192.5°$) for $\beta$-strands, and ($\theta = 92.5°$ and $\mu = 242.5°$) for loops.

We alert the reader that this grouping is distinct from the more familiar groupings of amino acids based on their non-local interactions (22-29). Here, instead, it is entirely based on the similarity of their propensity to adopt specific local conformations.

We defined three significantly occupied regions of ($\theta$, $\mu$) space corresponding to $\alpha$-helix ($\theta \in [90°,95°]$, $\mu \in [45°,50°]$), β-strand ($\theta \in [105°,145°]$, $\mu \in [175°,185°]$), and loop ($\theta \in [87.5°,97.5°]$, $\mu \in [90°,360°]$). The amino acid occupancies of the three regions are normalized by their frequencies in the entire ($\theta$, $\mu$) space of all 4416 proteins and they are shown in Table 2. Amino acids having a normalized occupancy greater than 1 are over-represented in a given region and vice versa compared to the expectation from random considerations. The over-represented amino acids in the $\alpha$-helix region (second column of Table 2) are all members of Groups A and C of amino acids with the top four being LEU (1.56), MET (1.46) and ALA/GLU both having 1.42 normalized occupancy. The amino acids over-represented in the β-strand region (third column of Table 2) are all members of amino acid Groups B and C, the top three being VAL (1.93), ILE (1.55) and TYR (1.51). Finally, the most over-represented amino acids in the loop region correspond to those that are the most under-represented in both the $\alpha$-helix and β-strand regions: PRO (2.49), GLY (1.76), ASP (1.33) and ASN (1.31). These four amino acids are members of the amino acid groups D (ASN and ASP), E (GLY), and F (PRO) – see amino acid grouping analysis and Figure 11. The strong correlation observed between the values of normalized occupancies of amino acids in the three regions and the results of the amino acid groupings suggests that amino acid Group A can be interpreted as the "$\alpha$-helical" group, amino acid Group B as the "β-strand" group, while Group C is over-represented in both $\alpha$-helix and β-strand regions. Finally, amino acid Groups D, E, and F can be described as "loop" groups, since they are strongly over-represented in loops and under-represented in both $\alpha$-helix and β-strand regions. These findings are in a good accord with the observed amino acid propensities in protens previously reported in the literature (3,5,31-33).

| Amino acid type | Normalized occupancy in the α-helix region | Normalized occupancy in the β-strand region | Normalized occupancy in the loop region |
|---|---|---|---|
| ALA | 1.42 | 0.85 | 0.89 |
| ARG | 1.29 | 1.02 | 0.89 |
| ASN | 0.77 | 0.60 | 1.31 |
| ASP | 0.77 | 0.48 | 1.33 |
| CYS | 0.87 | 1.41 | 0.67 |
| GLU | 1.42 | 0.63 | 0.97 |
| GLN | 1.38 | 0.78 | 0.88 |
| GLY | 0.34 | 0.60 | 1.76 |
| HIS | 0.80 | 0.98 | 0.81 |
| ILE | 1.26 | 1.55 | 0.51 |
| LEU | 1.56 | 0.94 | 0.70 |
| LYS | 1.21 | 0.75 | 1.08 |
| MET | 1.46 | 1.21 | 0.69 |
| PHE | 0.88 | 1.38 | 0.64 |
| PRO | 0.14 | 0.40 | 2.49 |
| SER | 0.61 | 1.05 | 1.04 |
| THR | 0.67 | 1.43 | 0.76 |
| TRP | 0.89 | 1.23 | 0.89 |
| TYR | 0.80 | 1.51 | 0.63 |
| VAL | 1.00 | 1.93 | 0.50 |

**Table 2: Propensity of the 20 amino acids to occupy the α-helix, β-strand, and loop regions in ($\theta$, $\mu$) space.** The numbers shown have been normalized by the amino acid occurrences in all of the ($\theta$, $\mu$) space.

With the identification of just six groups, we proceeded to an analysis of correlating the local structure ($\theta$, $\mu$) at bead i to the identity of the triplet of amino acid groups at positions (i-1,i,i+1). The simplicity now is that the total number of distinct triplets is 216 instead of 8000. We considered each of these triplets and studied the number of times these occurred. Obviously, one would expect that triplets containing the amino acids in groups C, D, E and F would be fewer than those occurring in Groups A and B. Indeed, the number of triplets which occurred more than 4461 times (deduced by dividing the total number of triplets = 963681 and the total number of types of triplets = 216) was just 57 and we used these for our analysis because of their statistical significance. The results are summarized in Figure 13.

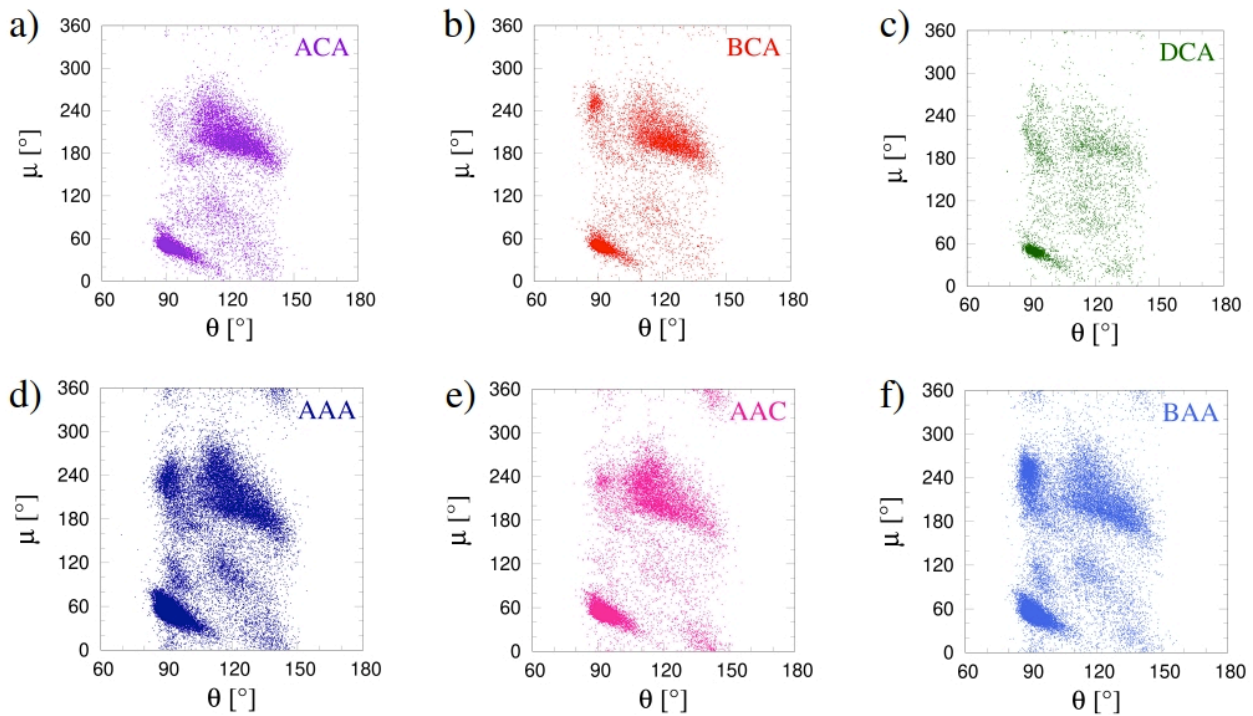

**Figure 13: The six panels show the distributions of the 6 most localized triplets in the (θ, μ) plane.** They all occupy the α-helix region predominantly. But they are spread out considerably underscoring the weak role of the amino acid sequence in matching with the local structure. We remind the reader (see Figure 3c) that the density peaks occur at (θ = 92.5° and μ = 47.5°) for α-helices, (θ = 122.5° and μ = 192.5°) for β-strands, and (θ = 92.5° and μ = 242.5°) for loops.

We conclude with the lessons learned from our analysis. Our goal here was to characterize the local structures associated with protein native state folds using the simple representation of just two angles (θ and μ) for each $C_\alpha$ position (Figure 1). This simplification is made possible because the vast majority of bond lengths is substantially constant (Figure 2). The (θ, μ) variables are a coarse-grained representation of successive Ramachandran angles. The local structures adopted by proteins are captured by simple patterns of points in the (θ, μ) plane. This reveals that protein native state structures (even at the local level) are highly structured unlike the behavior of a generic chain. Even though there is a great deal of spread in the θ and μ values, there is a tight correlation in the plot of the mean θ versus mean μ for the 4416 proteins (Figure 4).

Armed with insights on the local structural pattern, we explored a potential sequence-structure relationship in multiple ways. We considered the propensity of the 20 amino acids to occupy certain regions of local structural space. We also divided the 20 amino acids into 6 groups based on their similarity to each other in being associated with regions in the (θ, μ) space. We explored singlets and triplets based on grouping. The basic result of our analysis is that any sequence-local structure relationship is not very strong and there is flexibility in the ability of the amino acids to adapt to the local structure. This is consistent with the prevalence of neutral evolution where neither the native state fold nor the ability to function changes under many amino acid substitutions. It serves to underscore the pioneering results of Brian Matthews (34,35) and his team who "used the lysozyme from bacteriophage T4 to define the contributions that different types of interaction make to the stability of proteins". One of their key findings was that "the protein is, in general, very tolerant of amino acid replacement". Our findings also are in accord with more recent experimental studies on proteins (36,37) which showed that, while protein structures are highly tolerant of amino acid substitutions, a small number of key alterations can yield distinct structure and function. An interesting challenge is to be able to predict, in a transparent and reliable manner, the identity of these key amino acids.

We conclude by revisiting a seminal paper by Levitt (38) more than four decades ago in which he very carefully measured the Chou-Fasman propensity (39) of the twenty amino acids to be housed in three secondary structures. He noted that, generally, the preferences of the individual amino acids for secondary structure are rather weak. He provided a physical interpretation of his results by noting that "the chemical structure and sterochemistry of the amino acid plays a major part in determining its preference and dislike for secondary structure….. Bulky amino acids, namely, those that are branched at the β-carbon or have a large aromatic side chain, prefer β-sheet. The shorter polar side chains prefer reverse turns, as do Gly and Pro, the special side chains. All other side chains prefer α-helix, except Arg which has no preference." Table 3 shows a side-by-side comparison of the results of Levitt obtained with less than a hundred protein structures and our findings with entirely different methods and more than 4000 protein structures. Our results match those of Levitt (38) confirming the adage – *old is gold*.

| α-helix propensity | | β-sheet propensity | | Loop propensity | |
|---|---|---|---|---|---|
| Our study | Levitt [38] | Our study | Levitt [38] | Our study | Levitt [38] |
| LEU (1.56) | MET (1.47) | VAL (1.93) | VAL (1.49) | PRO (2.49) | PRO (1.91) |
| MET (1.46) | GLU (1.44) | ILE (1.55) | ILE (1.45) | GLY (1.76) | GLY (1.64) |
| GLU (1.42) | LEU (1.30) | TYR (1.51) | PHE (1.32) | ASP (1.33) | ASP (1.41) |

**Table 3: Identities of three amino acids with the highest propensities to occupy the α-helix, β-strand, and loop regions in (θ, μ) space (taken from Table 2).** The Table also shows the winning amino acids from Levitt's analysis of 1978 (38). There is excellent accord between our results and those of Levitt. The key difference is the identity of one of the top three amino acids in the β-sheet propensity group. PHE scores third in Levitt's analysis with a normalized probability of 1.32 whereas PHE scores fifth in our analysis with a similar probability score of 1.38. TYR scores third in our study and fourth in Levitt's analysis.

**Acknowledgements:** We are indebted to George Rose for his collaboration and inspiration. We are grateful to Pete von Hippel for his warm hospitality and to him and Brian Matthews for stimulating conversations. We are very thankful to two anonymous reviewers for their constructive suggestions. **Funding:** This project received funding from the European Union's Horizon 2020 research and innovation program under the Marie Skłodowska-Curie Grant Agreement No 894784. The contents reflect only the authors' view and not the views of the European Commission. Support from the University of Oregon (through a Knight Chair to JRB), University of Padova through "Excellence Project 2018" of the Cariparo foundation (AM), MIUR PRIN-COFIN2017 *Soft Adaptive Networks* grant 2017Z55KCW and COST action CA17139 (AG) is gratefully acknowledged. The computer calculations were performed on the Talapas cluster at the University of Oregon. **Conflict of interest:** The authors declare they have no conflict of interest.

**Author Contributions:** TŠ and JRB conceived the ideas for the calculations. TŠ carried out the calculations. JRB wrote the paper. All authors participated in understanding the results and reviewing the manuscript.

Emails: tskrbic@uoregon.edu, amos.maritan@pd.infn.it, achille.giacometti@unive.it, banavar@uoregon.edu

**References**

1. Ramachandran, G. N. & Sasisekharan, V. Conformation of polypeptides and proteins. *Adv. Prot. Chem.* **23**, 283-438 (1968).
2. Fitzkee, N. C. & Rose, G. D. Steric restrictions in protein folding: an $\alpha$-helix cannot be followed by a contiguous β-strand. *Protein Sci.* **13**, 633-639 (2004).
3. Lesk, A. M. *Introduction to Protein Science: Architecture, function and genomics* (Oxford University Press, 2004).
4. Rose, G. D., Fleming, P. J., Banavar, J. R. & Maritan, A. A backbone-based theory of protein folding. *Proc. Natl. Acad. Sci. USA* **103**, 16623-16633 (2006).
5. Bahar, I., Jernigan R. L. & Dill, K. A. *Protein Actions* (Garland Science, Taylor & Francis Group, 2017).

6. Rose, G. D. Protein Folding – Seeing is Deceiving, *preprint.*

7. Škrbić, T., Maritan, A., Giacometti, A., Rose, G. D. & Banavar, J. R. Building blocks of protein structures – Physics meets Biology, *bioRxiv doi: 10.1101/2020.11.10.375105*

*8.* Rubin B., & Richardson J. S. The simple construction of protein alpha-carbon models. *Biopolymers* **11**, 2381-2385 (1972).

9. https://proteopedia.org/wiki/index.php/Byron%27s_Bender

10. Lovell, S. C., Word, J.M., Richardson, J. S. & Richardson, D. C. The penultimate rotamer library. *Proteins* **40**, 389-408 (2000).

11. http://kinemage.biochem.duke.edu/databases/top8000.php.

12. Wang, G. & Dunbrack, R. L. Jr. PISCES: a protein sequence culling server. *Bioinformatics* **19**, 1589-1591 (2003).

13. Matthews, B. W. How planar are peptide bonds? *Protein Sci.* **25**, 776-777 (2016).

14. Wang, F.& Landau, D. Efficient, Multiple-Range Random Walk Algorithm to Calculate the Density of States. *Phys. Rev. Lett*. **86**, 2050–2053 (2001).

15. Flory, P. J. *Statistical Mechanics of Chain Molecules* (Wiley & Sons, 1968).

16. Rackovsky, S. & Scheraga H. A. Differential Geometry and Polymer Conformation. *Macromolecules* **14**, 1259-1269 (1981).

17. Levitt, M. A Simplified Representation of Protein Conformations for Rapid Simulation of Protein Folding. *J. Mol. Biol.* **104**, 59-107 (1976).

18. Oldfield, T. J. & Hubbard, R. E. Analysis of $C_\alpha$ Geometry in Protein Structures. *Proteins* **18**, 324-337 (1994).

19. DeWitte, R. S. & Shakhnovich, E. I. Pseudodihedrals: Simplified protein backbone representation with knowledge-based anergy. *Prot. Sci.* **3**, 1570-1581 (1994).

20. Bahar, I., Kaplan, M. & Jernigan, R. L. Short-Range Conformational Energies, Secondary Structure Propensities, and Recognition of Correct Sequence-Structure Matches. *Proteins* **26**, 292-308 (1997).

21. Bhattacharyya, A. On a measure of divergence between two statistical populations defined by their probability distributions. *Bulletin of the Calcutta Mathematical Society* **35**, 99–109 (1943).

22. Regan, L. & DeGrado, W. E. Characterization of a helical protein designed from first principles. *Science* **241**, 976-978 (1988).

23. Miyazawa S. & Jernigan, R. L. Residue-residue potentials with favorable contact pair-term and an unfavorable high packing density term, for simulation and threading. *J. Mol. Biol.* **256**, 623-644 (1996).

24. Riddle, D. S., Santiago, J. V., Bray-Hall, S. T., Doshi, N., Grantcharova, V. P., Yi, Q. & Baker D. Functional rapidly folding proteins from simplified amino acid sequences. *Nature Struct. Biol.* **4**, 805-809 (1997).

25  Wolynes, P. G. As simple as can be? *Nat. Struct. Biol.* **4**, 871-874 (1997).

26. Li, H., Tan, C. & Wingreen, N. S. Nature of Driving Force for Protein Folding: A Result From Analyzing the Statistical Potential. *Phys. Rev. Lett.* **79**, 765-768 (1997).

27. Wang J. & Wang, W. A computational approach to simplifying the protein folding alphabet. *Nat. Struct. Biol.* **6**, 1033-1038 (1999).

28. Chan, H. S. Folding alphabets. *Nat. Struct. Biol.* **6**, 994-996 (1999).

29. Cieplak, M., Holter, N. S., Maritan, A. & Banavar, J. R. Amino acid classes and the protein folding problem. *J. Chem. Phys.* **114**, 1420-1423 (2001).

30. Rose, G. D., Gierasch, L. M. & Smith, J. A. Turns in peptides and proteins. *Adv. Protein Chem.* **37**, 1-109 (1985).

31. Rose, G. D. & Presta, L. G. Helical signals in proteins. *Science* **240**, 1632-1641 (1988).

32. Petuhov, M. Uegaki, K., Yumoto, N. & Serrano L. Amino acid intrinsic α-helical propensities III: Positional dependence at several positions of C-terminus. Prot. Sci. **11**, 766-777 (2002).

33. Bhattacharjee, N. & Biswas, P. Position-specific propensities of amino acids in the β-strand -strand. *BMC Struct. Biol.* 10, **29** (2010).

34. Alber T., Bell J. A., Dao-Pin S., Nicholson H., Wozniak J. A., Cook S. & Matthews B. W. Replacements of Pro-86 in Phage T4 Lysozyme Extend an α-Helix But Do Not Alter Protein Stability. *Science* **239**, 631-635 (1988).

35. Matthews, B. W. Structural and genetic analysis of protein stability. *Annu. Rev. Biochem.* **62**, 139–160 (1993).

36. Alexander, P. A., He Y., Chen Y., Orban J. & Bryan P. N. The design and characterization of two proteins with 88% sequence identity but different structure and function. *Proc. Natl. Acad. Sci. USA* **104**, 11963–11968 (2007).

37. Alexander, P. A., He Y., Chen Y., Orban J. & Bryan P. N. A minimal sequence code for switching protein structure and function. *Proc. Natl. Acad. Sci. USA* **106**, 21149–21154 (2009).

38. Levitt, M. Conformational Preferences of Amino Acids in Globular Proteins. *Biochemistry* **17**, 4277-4285 (1978).

39. Chou, P. Y. & Fasman, G. D. Prediction of Protein Conformation. *Biochemistry* **13**, 222-245 (1974).

# Supplementary Information for the manuscript

# "Local sequence-structure relationships in proteins"

Tatjana Škrbić, Amos Maritan, Achille Giacometti and Jayanth R. Banavar

**Table S1: PDB codes of the 4416 proteins used in our analysis.**

| | | | | | | | | | |
|---|---|---|---|---|---|---|---|---|---|
| 16pk_A | 1iqc_C | 1pnc_A | 1w0p_A | 2buw_B | 2hwn_D | 2rc3_A | 2zk9_X | 3euf_D | 3kl0_B |
| 1a1i_A | 1iqq_A | 1pnd_A | 1w0u_A | 2bv2_B | 2hxm_A | 2rc8_B | 2zkd_B | 3eul_A | 3kl6_B |
| 1a2p_B | 1iqz_A | 1pp0_C | 1w1h_C | 2bv4_A | 2hxp_A | 2rci_A | 2zl6_B | 3eun_A | 3klq_A |
| 1a2y_A | 1irq_A | 1psr_A | 1w2c_A | 2bw0_A | 2hxs_A | 2rcq_A | 2znd_A | 3eup_B | 3klr_A |
| 1a2y_B | 1isp_A | 1ptq_A | 1w2i_B | 2bw8_A | 2hxt_A | 2rcv_E | 2znr_A | 3evf_A | 3kmt_C |
| 1a2z_C | 1isu_A | 1puc_A | 1w3i_A | 2bwf_A | 2hy5_A | 2rcz_B | 2zoo_A | 3evk_D | 3kmv_D |
| 1a34_A | 1it2_B | 1puf_B | 1w3w_A | 2bwl_A | 2hy5_B | 2rdh_C | 2zpd_A | 3evy_B | 3knb_B |
| 1a3a_A | 1itw_D | 1pvm_A | 1w3y_A | 2bwr_B | 2hy7_A | 2rdq_A | 2zpo_A | 3ew0_A | 3knv_A |
| 1a4i_B | 1itx_A | 1pvx_A | 1w4s_A | 2c0c_A | 2hyk_A | 2rdu_A | 2zpu_A | 3ew1_D | 3kp8_A |
| 1a73_A | 1iu8_B | 1pxv_B | 1w4t_A | 2c0h_A | 2hyv_A | 2rdz_A | 2zqe_A | 3ewi_A | 3kpb_D |
| 1a7d_A | 1iue_B | 1pyo_B | 1w4v_B | 2c0r_B | 2hzl_B | 2ree_A | 2zqm_A | 3exe_D | 3kq0_A |
| 1a7t_B | 1iuz_A | 1pzs_A | 1w4x_A | 2c0z_A | 2hzy_B | 2reg_A | 2zqn_B | 3exr_A | 3kqi_A |
| 1a88_A | 1iv3_D | 1q08_B | 1w53_A | 2c1d_D | 2i0q_A | 2rem_B | 2zs0_A | 3ey6_A | 3kqr_A |
| 1a8q_A | 1iv9_A | 1q0q_A | 1w5r_B | 2c1s_A | 2i1n_A | 2rer_A | 2zs0_D | 3eye_A | 3kre_A |
| 1a8s_A | 1iwd_A | 1q0r_A | 1w66_A | 2c1v_B | 2i24_N | 2rfg_A | 2zs1_B | 3eyi_A | 3krs_A |
| 1a92_C | 1ix1_B | 1q1r_B | 1w6s_C | 2c29_F | 2i2q_A | 2rfm_B | 2zs1_C | 3eyp_B | 3kru_A |
| 1ab1_A | 1ixg_A | 1q1u_A | 1w6s_D | 2c2n_A | 2i3f_A | 2rh2_A | 2zsi_A | 3ezi_B | 3kse_D |
| 1aba_A | 1iy8_C | 1q2h_A | 1w70_A | 2c2p_A | 2i49_A | 2rh3_A | 2ztl_C | 3f0y_C | 3ksh_A |
| 1afb_3 | 1iyb_A | 1q4u_B | 1w8o_A | 2c2u_A | 2i4a_A | 2rhi_A | 2zu1_B | 3f17_A | 3ksv_A |
| 1ag9_B | 1iye_C | 1q5m_B | 1w8u_A | 2c3n_C | 2i5r_B | 2rhk_C | 2zu2_A | 3f1l_A | 3ksx_A |
| 1agy_A | 1iyn_A | 1q5z_A | 1w99_A | 2c41_F | 2i5v_O | 2ri0_B | 2zux_B | 3f1p_A | 3kt9_A |
| 1ah7_A | 1izc_A | 1q6o_A | 1w9p_A | 2c42_B | 2i61_A | 2ri7_A | 2zuy_A | 3f1p_B | 3ktz_A |
| 1aho_A | 1ize_A | 1q7l_A | 1w9s_A | 2c4e_A | 2i62_D | 2ri9_A | 2zw2_A | 3f2e_A | 3ku3_B |
| 1aii_A | 1j05_B | 1q7l_B | 1wa3_A | 2c4f_T | 2i6v_A | 2rik_A | 2zwd_A | 3f2u_A | 3kus_B |
| 1ako_A | 1j0h_B | 1q8f_A | 1wb0_A | 2c4j_D | 2i7c_C | 2riq_A | 2zwj_A | 3f3q_A | 3kuv_A |
| 1aky_A | 1j0p_A | 1qau_A | 1wb6_B | 2c4n_A | 2i7d_A | 2rji_A | 2zwn_A | 3f3x_A | 3kwe_A |
| 1aoh_B | 1j1y_A | 1qav_A | 1wba_A | 2c53_A | 2i7f_B | 2rjw_A | 2zwu_A | 3f47_A | 3kxt_A |
| 1aoz_A | 1j24_A | 1qaz_A | 1wbe_A | 2c6q_B | 2i8t_B | 2rk3_A | 2zx2_A | 3f4m_A | 3kyj_A |
| 1arb_A | 1j27_A | 1qb5_E | 1wbh_B | 2c6u_A | 2i9a_D | 2rk5_A | 2zxj_B | 3f4s_A | 3kz5_A |
| 1ast_A | 1j2j_B | 1qb7_A | 1wbi_H | 2c6z_A | 2i9i_A | 2rkl_A | 2zxy_A | 3f52_A | 3kz7_A |
| 1atg_A | 1j2r_A | 1qba_A | 1wbj_A | 2c78_A | 2iax_A | 2rkq_A | 2zya_B | 3f5l_B | 3kzj_A |
| 1atl_B | 1j30_B | 1qcx_A | 1wbj_B | 2c7p_A | 2ib8_A | 2rku_A | 2zyh_B | 3f5o_G | 3kzu_B |
| 1atz_B | 1j34_A | 1qd1_B | 1wc2_A | 2c81_A | 2ibj_A | 2rky_C | 2zyo_A | 3f6o_A | 3l07_B |
| 1aun_A | 1j34_B | 1qd2_A | 1wc9_A | 2c82_B | 2ibl_A | 2sak_A | 2zzd_E | 3f6q_A | 3l0f_A |
| 1avb_A | 1j3w_C | 1qd9_C | 1wcf_A | 2c8h_D | 2ibp_B | 2sec_I | 2zzd_J | 3f6q_B | 3l0l_B |
| 1awd_A | 1j48_A | 1qdd_A | 1wcg_B | 2c92_D | 2ic6_A | 2sga_A | 2zzj_A | 3f6y_A | 3l18_A |
| 1aye_A | 1j71_A | 1qfv_B | 1wck_A | 2c95_B | 2ic7_B | 2sn3_A | 2zzr_A | 3f74_B | 3l1e_A |
| 1b0b_A | 1j75_A | 1qgi_A | 1wd3_A | 2c9q_A | 2idl_B | 2tnf_B | 2zzs_O | 3f75_A | 3l2c_A |
| 1b16_A | 1j77_A | 1qgj_A | 1wdd_S | 2cal_A | 2if6_A | 2uuy_B | 2zzv_B | 3f75_P | 3l32_A |
| 1b1c_A | 1j7d_A | 1qgu_D | 1wdy_A | 2car_B | 2ifc_C | 2uv4_A | 3a02_A | 3f7l_A | 3l39_A |
| 1b2s_F | 1j7g_A | 1qh5_B | 1wf3_A | 2cb5_A | 2ig8_A | 2uvj_A | 3a03_A | 3f7q_A | 3l3u_A |
| 1b37_B | 1j8e_A | 1qhf_A | 1whi_A | 2cb8_A | 2igi_A | 2uvo_B | 3a04_A | 3f8m_B | 3l41_A |
| 1b3a_B | 1j8u_A | 1qho_A | 1wka_A | 2cbz_A | 2igp_A | 2uw1_A | 3a07_A | 3f97_A | 3l42_A |
| 1b4f_B | 1j9l_A | 1qhq_A | 1wko_A | 2cc6_A | 2igv_A | 2uwa_A | 3a09_A | 3f9b_A | 3l46_A |
| 1b5e_A | 1ja9_A | 1qhv_A | 1wkq_B | 2cch_B | 2igx_A | 2uyt_A | 3a0y_B | 3f9r_A | 3l4p_A |
| 1b63_A | 1jae_A | 1qj5_B | 1wkr_A | 2ccq_A | 2ih5_A | 2uyw_A | 3a16_C | 3fas_B | 3l4r_A |
| 1b66_A | 1jak_A | 1qjc_B | 1wku_B | 2ccw_A | 2ihd_A | 2uyz_A | 3a1c_A | 3fb9_A | 3l5l_A |
| 1b67_A | 1jat_A | 1qjw_B | 1wkx_A | 2cdn_A | 2ii2_A | 2uyz_B | 3a21_A | 3fbg_A | 3l6g_A |
| 1b8a_B | 1jay_B | 1qkk_A | 1wld_A | 2cf7_C | 2iid_A | 2uz1_D | 3a2q_A | 3fbl_A | 3l6n_A |
| 1b8d_K | 1jcd_A | 1ql0_B | 1wlg_B | 2cfe_A | 2ijh_A | 2uzc_C | 3a2v_I | 3fd7_B | 3l77_A |
| 1b8p_A | 1jcv_A | 1ql3_B | 1wlz_C | 2cg7_A | 2ijq_A | 2v09_A | 3a2z_A | 3fde_B | 3l7h_B |
| 1b93_A | 1jd0_B | 1qlw_A | 1wm2_A | 2cgq_A | 2ijx_D | 2v0h_A | 3a39_A | 3fdl_A | 3l7t_B |
| 1bas_A | 1jd1_C | 1qmy_C | 1wma_A | 2chc_B | 2imf_A | 2v0s_A | 3a3d_B | 3fdq_A | 3l8e_B |
| 1baz_A | 1jd5_A | 1qnj_A | 1wmd_A | 2cia_A | 2imi_B | 2v0u_A | 3a3v_A | 3fdr_A | 3l8w_A |
| 1bdo_A | 1jdh_B | 1qnn_C | 1wmh_A | 2ciu_A | 2imq_X | 2v1o_B | 3a40_X | 3fe0_A | 3l91_A |
| 1beh_A | 1jdl_A | 1qnp_A | 1wmw_A | 2ciw_A | 2in8_A | 2v1q_A | 3a4r_A | 3fe7_A | 3l91_B |
| 1bf6_A | 1jek_A | 1qnx_A | 1wmz_D | 2cj3_A | 2inc_A | 2v1w_B | 3a4u_A | 3fev_A | 3l9a_X |
| 1bgf_A | 1jev_A | 1qoz_B | 1wn2_A | 2cj4_A | 2inc_B | 2v25_A | 3a4w_B | 3ff5_B | 3l9f_D |
| 1bgp_A | 1jf8_A | 1qre_A | 1wny_A | 2cjj_A | 2ior_A | 2v27_A | 3a57_A | 3ff7_C | 3l9s_A |
| 1bhp_A | 1jfl_B | 1qrp_E | 1wo8_D | 2cjl_B | 2ioy_B | 2v2g_C | 3a5p_D | 3ff9_B | 3l9u_A |
| 1bj7_A | 1jfr_A | 1qs1_A | 1wod_A | 2cjp_A | 2ip2_B | 2v33_B | 3a5r_A | 3fg0_F | 3l9y_B |
| 1bkp_A | 1jfu_A | 1qsa_A | 1wog_E | 2cjs_C | 2ip6_A | 2v36_D | 3a6r_B | 3fgd_A | 3las_B |
| 1bn8_A | 1jfx_A | 1qsg_A | 1woq_B | 2ckf_D | 2ipr_B | 2v3g_A | 3a72_A | 3fh2_A | 3lat_A |
| 1bq8_A | 1jg1_A | 1qt9_A | 1wor_A | 2ckk_A | 2iq7_A | 2v3s_A | 3a7l_A | 3fhg_A | 3lbe_D |
| 1bqb_A | 1jhd_A | 1qtn_A | 1wpa_A | 2cks_A | 2iqj_A | 2v4n_A | 3a7n_A | 3fid_A | 3lbf_C |
| 1bqk_A | 1jhf_A | 1qtw_A | 1wpn_B | 2cm4_A | 2iru_B | 2v4v_A | 3a8g_B | 3fil_B | 3lbl_A |
| 1brt_A | 1jhg_A | 1qu1_D | 1wpu_A | 2cmj_B | 2is8_A | 2v5i_A | 3a8u_X | 3fiq_A | 3lbm_B |
| 1bs3_B | 1jhj_A | 1qve_B | 1wq8_A | 2cmt_A | 2is9_A | 2v5j_A | 3a9b_A | 3fju_B | 3lcc_A |
| 1bs9_A | 1jhs_A | 1qw9_B | 1wqj_B | 2cn3_B | 2it1_A | 2v5z_A | 3a9f_A | 3fkb_E | 3lcm_A |
| 1bsg_A | 1ji1_A | 1qwd_A | 1wqj_I | 2cnz_A | 2iu5_A | 2v6a_O | 3a9j_A | 3fkc_A | 3ld3_A |
| 1bue_A | 1jid_A | 1qwg_A | 1wr8_B | 2cov_I | 2ium_A | 2v6k_B | 3a9l_B | 3fke_A | 3ldd_A |
| 1bx4_A | 1jif_A | 1qwk_A | 1wrd_A | 2cs7_A | 2ivf_A | 2v6u_A | 3a9q_N | 3flg_A | 3le0_A |
| 1bx7_A | 1jke_C | 1qwm_B | 1wri_A | 2cu5_A | 2ivf_B | 2v7w_C | 3a9s_B | 3flv_A | 3le3_A |
| 1bxu_A | 1jkg_A | 1qwz_A | 1wrm_A | 2cvd_D | 2ivn_A | 2v84_A | 3aa0_A | 3fn5_A | 3le4_A |
| 1bxy_A | 1jkx_A | 1qxy_A | 1ws8_A | 2cve_A | 2ivx_A | 2v89_A | 3aa6_B | 3fp5_A | 3let_A |
| 1byi_A | 1jl1_A | 1qy6_A | 1wst_A | 2cvi_B | 2ivy_A | 2v8i_A | 3aaf_B | 3fpc_A | 3lf6_B |
| 1c02_A | 1jl7_A | 1qz9_A | 1wt6_A | 2cwd_A | 2iw0_A | 2v8u_A | 3aal_A | 3fpf_A | 3lfh_F |
| 1c0p_A | 1jlj_A | 1r0r_E | 1wta_A | 2cwi_B | 2iw1_A | 2v9m_A | 3aam_A | 3fpk_B | 3lfj_B |
| 1c1d_A | 1jlt_A | 1r12_A | 1wte_A | 2cwr_A | 2iw2_B | 2v9t_B | 3ab6_A | 3fpr_D | 3lfk_C |
| 1c1k_A | 1jlt_B | 1r17_A | 1wtj_A | 2cws_A | 2iwk_A | 2v9v_A | 3aba_A | 3fpu_B | 3lg5_A |
| 1c1l_A | 1jm1_A | 1r1p_A | 1wto_A | 2cxn_B | 2iwz_A | 2vac_A | 3abf_E | 3fpw_A | 3lgi_A |
| 1c1y_A | 1jnr_C | 1r1t_B | 1wu9_B | 2cyg_A | 2ix4_B | 2vap_A | 3aci_A | 3fq3_C | 3lgn_A |
| 1c1y_B | 1jnr_D | 1r26_A | 1wui_S | 2cz4_A | 2ixc_A | 2vb1_A | 3act_B | 3fqm_A | 3lhq_A |
| 1c4q_B | 1jo0_A | 1r29_A | 1wur_B | 2czd_A | 2ixd_B | 2vba_D | 3acx_A | 3frq_A | 3lhr_B |
| 1c52_A | 1jo8_A | 1r2m_A | 1wve_D | 2czq_B | 2ixk_A | 2vbk_A | 3adg_A | 3frr_A | 3lid_B |
| 1c5e_A | 1jpe_A | 1r2r_B | 1wvf_A | 2d0i_B | 2ixm_A | 2vc3_A | 3ado_A | 3fs7_A | 3lim_D |
| 1c75_A | 1jq5_A | 1r3q_A | 1wwz_B | 2d16_A | 2izz_B | 2vc8_A | 3aey_A | 3ft1_C | 3liy_A |
| 1c7j_A | 1jqe_A | 1r45_B | 1wy1_A | 2d1c_A | 2j1s_A | 2ve8_E | 3afm_A | 3ftd_A | 3ljw_B |
| 1c7k_A | 1jr8_A | 1r55_A | 1wy2_B | 2d1x_A | 2j23_A | 2veb_A | 3afv_A | 3fv3_G | 3lke_A |
| 1c7n_F | 1jsd_B | 1r6d_A | 1wyx_B | 2d1y_C | 2j27_A | 2vfk_A | 3ag3_C | 3fv9_G | 3lkt_B |
| 1cc8_A | 1jt2_A | 1r6j_A | 1wz3_A | 2d29_A | 2j2j_F | 2vfq_A | 3ag3_E | 3fvb_B | 3lkt_Q |
| 1ccw_B | 1ju2_A | 1r6x_A | 1wz8_A | 2d37_A | 2j3x_A | 2vg1_B | 3ag7_A | 3fvh_A | 3llb_A |
| 1cf3_A | 1jub_B | 1r77_B | 1wzd_B | 2d3d_A | 2j5g_A | 2vg3_C | 3agn_A | 3fwa_A | 3llu_A |
| 1cg5_A | 1juv_A | 1r7j_A | 1x0c_A | 2d3n_A | 2j5i_F | 2vgp_D | 3ah2_A | 3fwy_A | 3lny_A |
| 1cg5_B | 1jvw_A | 1r87_A | 1x0l_A | 2d4n_A | 2j5y_A | 2vha_B | 3ahc_A | 3fx4_A | 3log_C |
| 1chd_A | 1jwq_A | 1r88_B | 1x1i_A | 2d4p_A | 2j5z_C | 2vi8_A | 3ahn_A | 3fx7_A | 3lp6_C |
| 1cip_A | 1jy2_N | 1r89_A | 1x1n_A | 2d4v_C | 2j6a_A | 2vig_A | 3ahx_D | 3fy1_B | 3lpc_A |
| 1cjc_A | 1jy2_R | 1r8h_D | 1x1o_B | 2d5b_A | 2j6b_A | 2vj0_A | 3ahy_A | 3fy3_A | 3lpe_B |
| 1cjw_A | 1jy3_P | 1r8s_A | 1x2i_A | 2d5c_A | 2j6f_A | 2vjp_B | 3ahz_A | 3fym_A | 3lpe_G |
| 1cka_A | 1jyh_A | 1r9d_A | 1x2t_C | 2d5k_C | 2j6i_A | 2vjv_B | 3ai3_C | 3fza_A | 3lpw_B |
| 1clc_A | 1jyo_B | 1r9h_A | 1x38_A | 2d5w_B | 2j6l_F | 2vk8_C | 3aia_A | 3fzy_B | 3lqw_A |
| 1cnv_A | 1k07_A | 1r9l_A | 1x3o_A | 2d5z_B | 2j73_B | 2vkj_A | 3aj7_A | 3g00_A | 3lr4_A |
| 1cnz_B | 1k0i_A | 1ra0_A | 1x3x_B | 2d68_B | 2j7j_A | 2vkl_A | 3ajo_A | 3g0e_A | 3lrt_A |
| 1coj_A | 1k0m_A | 1rc9_A | 1x46_A | 2d69_A | 2j7z_A | 2vkv_A | 3ajx_C | 3g0m_A | 3ls0_A |
| 1cpq_A | 1k1e_K | 1rdo_2 | 1x54_A | 2d6m_A | 2j8b_A | 2vla_A | 3ak2_B | 3g1l_A | 3ls9_A |
| 1cqm_B | 1k20_A | 1rfs_A | 1x6i_A | 2d73_A | 2j8g_A | 2vlf_B | 3ak8_H | 3g1p_B | 3ltj_A |
| 1cru_B | 1k2e_A | 1rfy_B | 1x6x_X | 2d7t_H | 2j8h_A | 2vlu_A | 3akb_A | 3g1v_A | 3luc_A |
| 1cs6_A | 1k4i_A | 1rg8_B | 1x8d_C | 2d7t_L | 2j8k_A | 2vmc_A | 3akh_A | 3g20_B | 3lum_D |
| 1ctj_A | 1k4m_C | 1rgx_C | 1x91_A | 2d81_A | 2j8m_A | 2vn4_A | 3alf_A | 3g21_A | 3lvf_P |
| 1cuo_A | 1k5c_A | 1rgz_A | 1x9i_A | 2d8d_B | 2j8w_A | 2vn6_A | 3alu_A | 3g2b_A | 3lw6_A |
| 1cxy_A | 1k5n_A | 1rh6_B | 1x9u_A | 2dc1_B | 2j9c_B | 2vn6_B | 3amn_B | 3g2s_B | 3lwg_B |
| 1cyd_D | 1k66_B | 1rh9_A | 1xcr_A | 2dc3_A | 2j9o_B | 2vng_A | 3ans_B | 3g46_B | 3lwz_A |
| 1cyo_A | 1k6a_A | 1rhc_A | 1xd3_C | 2dc4_B | 2j9w_A | 2vnk_C | 3ap9_A | 3g48_A | 3lx3_A |
| 1cz9_A | 1k6d_B | 1rie_A | 1xdw_A | 2ddb_C | 2ja2_A | 2vnl_A | 3apa_A | 3g5k_D | 3lxr_A |
| 1cza_N | 1k7c_A | 1rjd_A | 1xeo_A | 2ddx_A | 2jab_C | 2vnz_X | 3apr_E | 3g5w_C | 3lxr_F |
| 1czf_A | 1k7i_A | 1rkd_A | 1xes_B | 2de3_A | 2jaf_A | 2vo4_A | 3b34_A | 3g6m_A | 3lxy_A |
| 1czn_A | 1k94_A | 1rki_A | 1xfk_A | 2de6_A | 2jb0_A | 2vo8_A | 3b42_A | 3g7n_B | 3ly0_B |
| 1d02_B | 1k9u_B | 1rkq_A | 1xg0_B | 2de6_F | 2jba_B | 2vo9_B | 3b4u_B | 3g7w_A | 3ly7_A |
| 1d0d_A | 1ka1_A | 1rku_A | 1xg2_A | 2dep_A | 2jbv_A | 2voc_A | 3b4w_A | 3g8h_A | 3lz5_A |
| 1d2n_A | 1kaf_A | 1rl0_A | 1xg2_B | 2dfb_A | 2jc4_A | 2voz_A | 3b51_X | 3g98_B | 3lzo_B |
| 1d4o_A | 1kao_A | 1rlh_A | 1xg4_A | 2dfd_C | 2jc5_A | 2vpg_A | 3b5g_B | 3g9m_B | 3lzw_A |
| 1d4t_A | 1kap_P | 1rlk_A | 1xg7_B | 2dg1_C | 2jcb_A | 2vpj_A | 3b5l_B | 3g9x_A | 3m07_A |
| 1d5l_B | 1kaz_A | 1rm6_A | 1xgs_A | 2dga_A | 2jcq_A | 2vq4_A | 3b5m_B | 3ga3_A | 3m0f_A |
| 1d5t_A | 1kdg_B | 1rm6_B | 1xiw_A | 2dge_B | 2jda_A | 2vqg_D | 3b5n_A | 3ga4_A | 3m0j_A |
| 1d7o_A | 1kdj_A | 1rm6_C | 1xiw_C | 2dgk_A | 2jdd_A | 2vri_A | 3b5n_B | 3gad_F | 3m0m_B |
| 1daa_B | 1kdo_B | 1roc_A | 1xiw_H | 2dho_A | 2jdf_A | 2vrs_C | 3b5n_D | 3gah_A | 3m1h_B |
| 1dbf_A | 1keq_B | 1rp0_B | 1xk4_H | 2dkj_A | 2jdk_D | 2vsv_A | 3b5n_K | 3gbe_A | 3m21_D |
| 1dbw_B | 1kew_A | 1rqj_A | 1xk4_I | 2dko_A | 2je6_B | 2vtc_B | 3b64_A | 3gbs_A | 3m3g_A |
| 1dci_C | 1kfw_A | 1rro_A | 1xky_B | 2dm9_B | 2je8_B | 2vuj_A | 3b6i_A | 3gc6_A | 3m4d_A |
| 1deu_A | 1kg2_A | 1rtq_A | 1xkz_B | 2dp6_A | 2jek_A | 2vun_B | 3b76_A | 3gcz_A | 3m5l_A |
| 1dfu_P | 1kgc_D | 1rtt_A | 1xlq_C | 2dp9_A | 2jep_B | 2vuo_A | 3b7e_A | 3gd6_A | 3m5q_A |
| 1dgf_A | 1khi_A | 1ru0_B | 1xm8_A | 2dpf_D | 2jft_A | 2vv6_D | 3b7s_A | 3gd8_A | 3m66_A |
| 1dhn_A | 1khq_A | 1ru4_A | 1xmk_A | 2dqa_A | 2jg6_A | 2vve_A | 3b84_A | 3gdc_A | 3m6b_A |
| 1dj0_B | 1kid_A | 1rv9_A | 1xmp_B | 2dql_A | 2jh1_A | 2vvp_B | 3b8f_C | 3gdl_B | 3m6z_A |
| 1djr_G | 1kjv_A | 1rwh_A | 1xmt_A | 2dr1_B | 2jhf_B | 2vvt_B | 3b8i_E | 3ge3_A | 3m73_A |
| 1dk8_A | 1klx_A | 1rwj_A | 1xng_B | 2dri_A | 2jhq_A | 2vvw_A | 3b8z_B | 3ge3_E | 3m7o_A |
| 1dl5_B | 1km9_A | 1rwr_A | 1xnk_A | 2drm_B | 2ji7_A | 2vw8_A | 3b9c_C | 3gfu_A | 3m7q_B |
| 1dlf_H | 1kms_A | 1rwy_B | 1xo7_B | 2ds2_D | 2jik_A | 2vwf_A | 3b9d_A | 3gg7_A | 3m8j_A |
| 1dlf_L | 1kmt_A | 1rwz_A | 1xoc_A | 2ds5_A | 2jjc_A | 2vwr_A | 3b9w_A | 3gg9_B | 3m8o_H |
| 1dlj_A | 1kng_A | 1rx0_C | 1xov_A | 2dsj_B | 2jjf_A | 2vws_C | 3ba1_A | 3ggw_C | 3m8o_L |
| 1dlw_A | 1knt_A | 1ry9_C | 1xph_A | 2dsn_B | 2jjn_A | 2vx5_A | 3baa_A | 3ggy_A | 3m8t_B |
| 1dly_A | 1koe_A | 1ryi_B | 1xpp_C | 2dsx_A | 2jjs_C | 2vxn_A | 3bal_B | 3gh6_A | 3m8u_A |
| 1dm1_A | 1kol_B | 1ryo_A | 1xqo_A | 2dt4_A | 2jk9_A | 2vxq_A | 3bbb_D | 3gha_A | 3m91_C |
| 1doi_A | 1kop_A | 1ryp_I | 1xre_A | 2dtj_A | 2jkb_A | 2vxt_I | 3bc1_B | 3gip_A | 3m9q_B |
| 1dok_A | 1kp6_A | 1ryp_J | 1xrk_B | 2dtx_A | 2jkg_A | 2vxt_L | 3bc9_A | 3gir_A | 3mab_A |
| 1dp7_P | 1kpt_A | 1ryp_K | 1xs5_A | 2dur_A | 2jkh_A | 2vxy_A | 3bd1_A | 3gk7_B | 3mao_A |
| 1dpj_A | 1kpu_B | 1ryp_Z | 1xso_B | 2dvm_B | 2jl1_A | 2vyo_A | 3bd2_A | 3gkb_C | 3maz_A |
| 1dpt_A | 1kq1_B | 1rzh_H | 1xsz_A | 2dvx_B | 2jlp_A | 2vyq_A | 3bec_A | 3gkj_A | 3mb4_B |
| 1dqg_A | 1kq3_A | 1rzh_L | 1xt5_A | 2dwu_A | 2jlq_A | 2vyw_A | 3beo_A | 3gkm_A | 3mb5_A |
| 1dqi_A | 1kqp_A | 1rzh_M | 1xtt_C | 2dxe_B | 2lis_A | 2vyx_C | 3ber_A | 3gkr_A | 3mbg_B |
| 1dqp_A | 1kqr_A | 1s1d_B | 1xty_B | 2dy0_A | 2mnr_A | 2vzc_A | 3bex_A | 3gkt_A | 3mbk_B |
| 1dqz_A | 1krh_A | 1s2o_A | 1xu1_D | 2dy1_A | 2nml_A | 2vzm_A | 3bf7_B | 3gkv_B | 3mbx_H |
| 1ds1_A | 1krn_A | 1s3c_A | 1xu1_T | 2dyj_B | 2nmx_A | 2w0p_B | 3bfk_B | 3gl0_A | 3mcb_B |
| 1dsx_C | 1ks8_A | 1s57_B | 1xvg_C | 2dyr_K | 2nn8_A | 2w15_A | 3bfo_B | 3glr_A | 3md1_A |
| 1dsz_A | 1ku1_A | 1s5m_A | 1xvg_E | 2dyu_A | 2nnr_A | 2w1j_B | 3bfq_G | 3glv_B | 3md7_A |
| 1dtd_B | 1kug_A | 1s5u_D | 1xvo_A | 2dze_A | 2nnu_A | 2w1p_A | 3bfv_A | 3gmf_A | 3md9_A |
| 1duv_H | 1kvd_C | 1s69_A | 1xvx_A | 2e0q_A | 2no4_A | 2w1r_A | 3bg8_A | 3gmg_A | 3mdm_A |
| 1dxe_B | 1kve_D | 1s9r_A | 1xw6_A | 2e0t_A | 2npt_D | 2w1v_A | 3bgo_P | 3gmi_A | 3mds_B |
| 1dxj_A | 1kw6_B | 1sa3_A | 1xwt_A | 2e11_B | 2nql_B | 2w20_A | 3bh4_A | 3gms_A | 3mdu_A |
| 1dxy_A | 1kwf_A | 1sat_A | 1xwv_A | 2e1n_B | 2ns1_B | 2w2b_B | 3bh7_A | 3gmv_X | 3meb_B |
| 1dys_B | 1kwg_A | 1sau_A | 1xww_A | 2e1z_A | 2nsf_A | 2w2j_A | 3bh7_B | 3gmx_B | 3mf7_A |
| 1dzk_A | 1kzq_A | 1sbp_A | 1xx1_C | 2e27_L | 2nsz_A | 2w2k_A | 3bj1_C | 3gn9_C | 3mgn_B |
| 1e0w_A | 1l2p_A | 1sd5_A | 1xxq_D | 2e2o_A | 2nt4_A | 2w31_B | 3bje_A | 3gne_A | 3mh9_A |
| 1e25_A | 1l2t_B | 1sdi_A | 1xyz_A | 2e3a_A | 2ntp_A | 2w39_A | 3bkb_A | 3gnr_A | 3mhs_B |
| 1e29_A | 1l3p_A | 1sds_A | 1y0h_A | 2e3z_B | 2nug_B | 2w3g_A | 3bkj_H | 3go2_A | 3mhy_C |
| 1e2w_B | 1l5o_A | 1seg_A | 1y0m_A | 2e42_A | 2nuh_A | 2w3j_A | 3bkr_A | 3go6_A | 3mi4_A |
| 1e3d_B | 1l5w_B | 1sen_A | 1y1p_A | 2e4t_A | 2nuk_A | 2w3p_B | 3bkt_A | 3goc_B | 3mil_A |
| 1e4c_P | 1l6r_A | 1sf9_A | 1y1x_A | 2e5f_A | 2nuw_A | 2w3v_A | 3bl6_A | 3goe_A | 3mjo_B |
| 1e4m_M | 1l6w_B | 1sff_C | 1y20_A | 2e5y_B | 2nw2_B | 2w3x_A | 3bmb_B | 3gon_A | 3mjv_B |
| 1e4v_V | 1l7l_A | 1sfs_A | 1y2t_B | 2e6f_A | 2nx0_A | 2w40_A | 3bmw_A | 3gox_B | 3mkh_B |
| 1e59_A | 1l9l_A | 1sg4_C | 1y37_A | 2e6u_X | 2nx4_C | 2w43_A | 3bmx_B | 3gp3_D | 3mlb_A |
| 1e5k_A | 1l9x_A | 1sgw_A | 1y43_B | 2e7u_A | 2nxb_B | 2w47_A | 3bn6_A | 3gp4_B | 3mm5_B |
| 1e5m_A | 1lb6_A | 1sh7_B | 1y4j_B | 2e7z_A | 2nyb_A | 2w4c_A | 3bnj_A | 3gpg_B | 3mm6_A |
| 1e6i_A | 1lc3_A | 1sh8_B | 1y4w_A | 2e85_A | 2nz7_A | 2w4f_A | 3bo6_B | 3gqh_A | 3mmg_A |
| 1e6y_B | 1lc5_A | 1shu_X | 1y51_A | 2e8e_A | 2nzh_A | 2w4i_F | 3bod_A | 3gqj_A | 3mmh_B |
| 1e7l_B | 1lcl_A | 1skz_A | 1y5i_B | 2e9m_A | 2o07_B | 2w50_B | 3boe_A | 3grh_A | 3mmw_D |
| 1e7s_A | 1lcp_A | 1smo_A | 1y60_E | 2e9y_B | 2o0b_A | 2w5q_A | 3boi_A | 3grn_A | 3mn1_C |
| 1e9g_A | 1le6_A | 1sn2_B | 1y66_C | 2ea3_A | 2o1c_B | 2w70_A | 3bom_C | 3gru_A | 3mos_A |
| 1eaj_B | 1lgt_A | 1snn_A | 1y6i_A | 2eab_A | 2o1k_B | 2w7n_A | 3bom_D | 3gsh_A | 3moy_A |
| 1eao_A | 1lj5_A | 1stm_B | 1y7p_B | 2ebb_A | 2o20_F | 2w7w_B | 3bov_A | 3gt5_A | 3mpc_A |
| 1ear_A | 1lj9_B | 1svb_A | 1y7t_B | 2ebo_C | 2o28_A | 2w83_C | 3bp5_A | 3gv6_A | 3mpz_B |
| 1eb6_A | 1ljo_A | 1svd_M | 1y7w_A | 2ecs_A | 2o2c_C | 2w86_A | 3bpj_B | 3gvf_A | 3mq2_B |
| 1eco_A | 1lk5_B | 1svf_B | 1y80_A | 2ecu_A | 2o2p_D | 2w8x_A | 3bpv_A | 3gvg_B | 3mqd_A |
| 1edg_A | 1lkp_A | 1sw5_B | 1y8a_A | 2eeo_B | 2o36_A | 2w8y_B | 3bpw_A | 3gvo_A | 3mqh_D |
| 1edq_A | 1llf_B | 1swy_A | 1y9l_A | 2eey_A | 2o37_A | 2w91_A | 3bpz_D | 3gw9_C | 3ms5_A |
| 1eej_B | 1llm_C | 1sxq_B | 1y9w_B | 2efr_B | 2o4a_A | 2w98_B | 3bqp_B | 3gwa_A | 3msu_B |
| 1egw_B | 1lo6_A | 1sxr_A | 1y9z_B | 2efv_A | 2o4t_A | 2w9h_A | 3br8_A | 3gwc_D | 3msx_B |
| 1eis_A | 1lq9_A | 1sy7_B | 1yac_B | 2egd_B | 2o4v_B | 2w9r_A | 3bs1_A | 3gwh_B | 3mte_A |
| 1ej0_A | 1lqa_A | 1syy_A | 1yar_D | 2egj_A | 2o5f_B | 2wa2_B | 3bs2_A | 3gwi_A | 3mtr_B |
| 1ej2_A | 1lqv_A | 1szh_A | 1yar_N | 2egv_A | 2o5g_A | 2waa_A | 3bs9_B | 3gwk_E | 3mu7_A |
| 1ejb_A | 1lqx_A | 1szn_A | 1yb0_B | 2eh6_A | 2o5u_A | 2wb6_A | 3bsy_B | 3gwm_A | 3muj_B |
| 1ekj_C | 1lr7_A | 1szo_K | 1yb5_A | 2ehg_A | 2o6f_A | 2wbf_X | 3buv_B | 3gwn_A | 3muz_3 |
| 1ekx_A | 1lrh_A | 1t00_A | 1ybi_B | 2ehq_A | 2o6p_A | 2wbs_A | 3bv6_D | 3gx8_A | 3mwc_A |
| 1elk_A | 1ls6_A | 1t06_A | 1ybk_D | 2eht_A | 2o6s_A | 2wc8_B | 3bvk_F | 3gxb_A | 3mwf_A |
| 1elr_A | 1ls9_A | 1t0b_D | 1ybz_A | 2ehz_A | 2o74_F | 2wci_A | 3bwh_A | 3gxr_B | 3mwj_A |
| 1elu_A | 1lst_A | 1t0f_B | 1yd3_A | 2ei5_B | 2o7i_A | 2wcj_A | 3bwu_D | 3gzg_A | 3mx6_B |
| 1elw_A | 1lt1_H | 1t0f_C | 1ydy_A | 2eiq_B | 2o90_A | 2wco_A | 3bx4_A | 3gzh_A | 3mxn_A |
| 1enf_A | 1lua_C | 1t0p_B | 1yfn_C | 2eix_A | 2o9c_A | 2wcr_A | 3bx4_B | 3gzk_A | 3mxn_B |
| 1eo6_B | 1lvw_B | 1t0t_X | 1yfu_A | 2eiy_B | 2o9s_A | 2wcu_A | 3bxe_B | 3gzx_A | 3mxu_A |
| 1ep0_A | 1lw6_E | 1t1g_A | 1yif_A | 2eja_B | 2oaa_B | 2wdc_A | 3by4_A | 3gzx_B | 3myb_A |
| 1eq9_B | 1lw6_I | 1t1j_B | 1yii_A | 2ejn_B | 2obi_A | 2wds_A | 3byb_B | 3h01_A | 3mzv_B |
| 1es5_A | 1lwb_A | 1t1v_B | 1ykd_A | 2ejw_A | 2obl_A | 2wdu_B | 3byp_A | 3h04_A | 3n08_A |
| 1esw_A | 1lwd_B | 1t2d_A | 1yki_B | 2ejx_A | 2ocg_A | 2we5_C | 3bzz_B | 3h09_B | 3n0i_B |
| 1eu3_A | 1ly2_A | 1t2h_B | 1yn3_B | 2ekp_A | 2ode_D | 2wei_A | 3c05_A | 3h0o_A | 3n10_B |
| 1euh_C | 1m0d_C | 1t2w_C | 1yn8_E | 2eky_C | 2odf_E | 2wf6_A | 3c05_D | 3h0u_C | 3n11_A |
| 1euv_A | 1m0s_B | 1t3q_A | 1yn9_A | 2elc_B | 2odk_C | 2wfc_C | 3c0i_A | 3h12_B | 3n1e_B |
| 1euv_B | 1m0u_A | 1t3q_B | 1ynb_C | 2end_A | 2oe3_A | 2wfh_B | 3c1o_A | 3h1g_A | 3n1f_D |
| 1evh_A | 1m15_A | 1t3y_A | 1ynh_B | 2eo4_A | 2oeb_A | 2wfj_A | 3c2u_A | 3h1s_B | 3n1s_M |
| 1ex2_B | 1m1n_E | 1t4b_B | 1ynp_B | 2eq6_B | 2ofc_A | 2wfo_A | 3c3y_B | 3h34_A | 3n22_A |
| 1ext_A | 1m1n_F | 1t61_C | 1yo3_A | 2erf_A | 2ofk_A | 2wfz_A | 3c4s_A | 3h3n_X | 3n2n_E |
| 1eyh_A | 1m1r_A | 1t61_E | 1yoa_A | 2erw_A | 2og1_A | 2wge_A | 3c5a_A | 3h4n_A | 3n37_A |
| 1eyl_A | 1m2d_B | 1t6c_A | 1yoc_B | 2ery_B | 2oh5_A | 2wgp_B | 3c5e_A | 3h4x_A | 3n3s_A |
| 1eyv_A | 1m2h_A | 1t6g_D | 1yon_A | 2esl_A | 2ohw_B | 2wh7_A | 3c5j_A | 3h55_B | 3n4i_B |
| 1ez3_B | 1m2t_B | 1t6u_L | 1yp1_A | 2et1_A | 2oif_B | 2whg_B | 3c5k_A | 3h5i_A | 3n4j_A |
| 1ezg_B | 1m2x_D | 1t6v_N | 1yph_D | 2etb_A | 2oiz_B | 2wi8_A | 3c68_B | 3h5j_B | 3n5a_A |
| 1ezm_A | 1m3u_C | 1t7q_B | 1yph_E | 2etx_A | 2oj6_C | 2wiy_A | 3c6w_A | 3h5l_B | 3n5b_B |
| 1f08_B | 1m40_A | 1t7r_A | 1yq2_C | 2eu7_X | 2okl_B | 2wj5_A | 3c70_A | 3h62_B | 3n72_B |
| 1f0k_B | 1m4i_A | 1t8h_A | 1yqd_A | 2eut_A | 2okm_A | 2wje_A | 3c7f_A | 3h6p_B | 3n79_A |
| 1f0y_B | 1m4j_A | 1t8k_A | 1yqe_A | 2ev1_A | 2okq_A | 2wjn_C | 3c7t_A | 3h6p_C | 3n98_A |
| 1f1m_C | 1m55_A | 1t8t_B | 1yqw_B | 2evb_A | 2ol1_B | 2wjn_L | 3c7x_A | 3h78_A | 3n9g_H |
| 1f1u_A | 1m5t_A | 1t8z_C | 1yqw_Q | 2ewh_A | 2olm_A | 2wjn_M | 3c8e_A | 3h7h_A | 3n9i_B |
| 1f39_A | 1m6j_B | 1t92_B | 1yqz_A | 2ewt_A | 2oln_A | 2wk0_A | 3c8i_A | 3h7h_B | 3n9u_B |
| 1f3u_G | 1m70_D | 1t9i_B | 1yrk_A | 2ex0_B | 2olp_A | 2wkk_C | 3c8o_A | 3h7i_A | 3n9u_C |
| 1f46_B | 1m7a_B | 1ta3_A | 1ys1_X | 2ex2_A | 2olr_A | 2wkx_A | 3c8p_A | 3h7r_A | 3nbk_A |
| 1f4p_A | 1m7g_A | 1ta3_B | 1ysl_B | 2exh_D | 2omy_B | 2wl1_A | 3c97_A | 3h7u_A | 3ncl_A |
| 1f5j_A | 1m7j_A | 1ta9_B | 1yt3_A | 2exv_A | 2omz_A | 2wm3_A | 3c9a_B | 3h81_C | 3ndd_A |
| 1f5v_A | 1m7s_D | 1tag_A | 1ytq_A | 2ez9_A | 2on5_A | 2wm8_A | 3c9h_B | 3h87_B | 3ndh_B |
| 1f60_A | 1m8s_A | 1taw_B | 1yu0_A | 2f0c_A | 2oo1_B | 2wmf_A | 3c9u_B | 3h87_D | 3ndj_A |
| 1f60_B | 1m93_B | 1tbf_A | 1yuz_B | 2f23_B | 2op3_A | 2wn3_C | 3c9x_A | 3h8g_C | 3ndo_A |
| 1f7l_A | 1m9z_A | 1tc5_B | 1yw5_A | 2f2b_A | 2op6_A | 2wnp_F | 3c9z_A | 3h8t_A | 3nfu_A |
| 1f8m_C | 1mb4_A | 1tca_A | 1ywm_A | 2f51_A | 2opc_A | 2wns_A | 3ca7_A | 3h8x_A | 3nfw_B |
| 1f94_A | 1mc2_A | 1ten_A | 1yxy_A | 2f5g_B | 2opg_B | 2wnv_F | 3cai_A | 3h91_A | 3ng7_X |
| 1f9f_D | 1md6_A | 1tez_B | 1yya_A | 2f5t_X | 2oqb_A | 2wnx_A | 3cb0_D | 3h9c_A | 3ngf_A |
| 1fcz_A | 1me4_A | 1tf1_A | 1yzf_A | 2f5v_A | 2or7_A | 2wny_B | 3cb6_A | 3h9e_O | 3ngh_A |
| 1fd3_A | 1mex_L | 1tf4_A | 1yzl_A | 2f60_K | 2os5_A | 2woe_C | 3cbq_A | 3ha9_A | 3ngj_A |
| 1fe0_B | 1mfa_H | 1tg0_A | 1yzm_A | 2f6e_A | 2os9_B | 2wol_A | 3cbx_A | 3hc4_L | 3ngp_A |
| 1fec_B | 1mg4_A | 1tg7_A | 1yzq_A | 2f6u_A | 2osa_A | 2wot_A | 3ccd_B | 3hcj_A | 3ni0_A |
| 1feh_A | 1mgq_C | 1tgr_A | 1yzx_A | 2f7b_A | 2otu_E | 2wp7_A | 3ce6_A | 3hcn_B | 3ni2_A |
| 1fj2_B | 1mgr_A | 1tgx_B | 1z06_A | 2f91_A | 2oui_A | 2wpq_C | 3ce7_A | 3hd4_A | 3nis_B |
| 1fk5_A | 1mgt_A | 1th7_H | 1z08_C | 2f99_C | 2ous_A | 2wpv_D | 3cfc_H | 3hdf_B | 3nj2_A |
| 1flj_A | 1mhn_A | 1thg_A | 1z0j_A | 2f9i_D | 2ov0_A | 2wpv_E | 3cfz_A | 3hdl_A | 3nje_B |
| 1fm0_D | 1mhx_A | 1thm_A | 1z0j_B | 2f9n_B | 2ow9_B | 2wqi_D | 3cg1_A | 3he5_C | 3njn_C |
| 1fm4_A | 1mi3_B | 1thx_A | 1z0s_C | 2fao_B | 2ows_A | 2wqk_A | 3cgi_C | 3he5_D | 3nn1_A |
| 1fn9_A | 1mix_A | 1thz_A | 1z1s_A | 2fb5_A | 2ox0_A | 2wqr_A | 3chj_A | 3he8_B | 3no0_A |
| 1fob_A | 1mj5_A | 1tjy_A | 1z2a_A | 2fba_A | 2ox4_H | 2wsb_C | 3chm_A | 3hef_B | 3no7_A |
| 1fp2_A | 1mk0_A | 1tke_A | 1z2n_X | 2fbd_A | 2ox6_B | 2wt1_A | 3ci7_A | 3hf5_C | 3noj_A |
| 1fpo_B | 1mkk_A | 1tn4_A | 1z2u_A | 2fbn_A | 2oxc_A | 2wta_A | 3cij_A | 3hfo_A | 3nok_A |
| 1fqt_B | 1mla_A | 1to4_A | 1z3e_A | 2fbq_A | 2oxg_A | 2wtg_A | 3cin_A | 3hfw_A | 3nol_A |
| 1fr3_A | 1mlw_A | 1toa_B | 1z3e_B | 2fc3_A | 2oxg_Y | 2wtm_C | 3cip_G | 3hg3_B | 3noo_B |
| 1frb_A | 1mn8_B | 1tov_A | 1z3q_A | 2fcb_A | 2oxn_A | 2wto_A | 3civ_A | 3hgb_A | 3nqx_A |
| 1fsg_A | 1mo0_B | 1tp5_A | 1z47_A | 2fcr_A | 2oy2_F | 2wu9_A | 3cjp_B | 3hgm_A | 3nr1_A |
| 1ftr_A | 1mo9_A | 1tp6_A | 1z4j_A | 2fcw_A | 2oy7_A | 2wue_A | 3cjs_B | 3hgu_A | 3ns2_A |
| 1fus_A | 1mof_A | 1tp9_B | 1z4r_A | 2fcw_B | 2oya_B | 2wuh_A | 3ckf_A | 3hh7_A | 3ns6_B |
| 1fw4_A | 1moq_A | 1tqh_A | 1z6n_A | 2fd5_A | 2oyn_A | 2wuk_A | 3cl5_A | 3hh8_A | 3nsl_F |
| 1fx4_A | 1mpg_B | 1tr0_J | 1z6o_D | 2fdn_A | 2oyp_A | 2wut_A | 3cla_A | 3hhi_B | 3nsx_B |
| 1fxl_A | 1mpl_A | 1tsf_A | 1z6o_M | 2fdv_A | 2ozf_A | 2wv3_A | 3cls_D | 3hht_B | 3nsz_A |
| 1fxo_G | 1mug_A | 1tt2_A | 1z76_B | 2fdx_A | 2ozl_A | 2wvf_A | 3cm0_A | 3his_A | 3nt1_B |
| 1fz1_B | 1mv8_C | 1tt8_A | 1z7x_W | 2fe3_A | 2ozn_A | 2wvg_F | 3cmj_A | 3hjb_A | 3ntk_A |
| 1g01_A | 1mvf_A | 1tu1_B | 1z8o_A | 2fe5_A | 2ozn_B | 2wvv_A | 3cmy_A | 3hje_A | 3nua_A |
| 1g0o_C | 1mvf_D | 1tu7_B | 1z96_A | 2fe8_A | 2p02_A | 2ww2_B | 3cnk_B | 3hjr_A | 3nv1_A |
| 1g12_A | 1mvo_A | 1tu9_A | 1zbf_A | 2ff4_A | 2p09_A | 2ww5_A | 3cnm_A | 3hjs_A | 3nvs_A |
| 1g1s_B | 1mxg_A | 1tua_A | 1zch_A | 2ffu_A | 2p0f_A | 2wwe_A | 3cnu_A | 3hl5_B | 3nvw_J |
| 1g1t_A | 1mxr_A | 1tuh_A | 1zcz_A | 2ffy_B | 2p1g_B | 2wwf_C | 3cou_A | 3hlx_A | 3nwo_A |
| 1g29_2 | 1my5_A | 1tuk_A | 1zd0_A | 2fgo_A | 2p1m_B | 2wwk_T | 3cp5_A | 3hm5_A | 3nxb_B |
| 1g2o_C | 1my6_B | 1tvn_B | 1zd8_A | 2fgr_A | 2p39_A | 2wwx_A | 3cp7_B | 3hmc_A | 3ny3_A |
| 1g2q_A | 1mz4_A | 1txg_B | 1zdy_A | 2fhf_A | 2p3h_A | 2wwx_B | 3cpq_B | 3hms_A | 3ny7_A |
| 1g2r_A | 1mz9_A | 1ty9_A | 1zem_A | 2fhz_A | 2p49_B | 2wx9_A | 3cpt_A | 3hnb_M | 3nye_A |
| 1g3k_B | 1mzy_A | 1tzw_A | 1zgd_B | 2fhz_B | 2p4e_P | 2wy3_B | 3cq5_B | 3hnx_A | 3nyk_A |
| 1g4i_A | 1n08_B | 1tzy_C | 1zhh_A | 2fi1_A | 2p4k_A | 2wy4_A | 3cql_A | 3hol_A | 3nyt_A |
| 1g5a_A | 1n0q_B | 1tzy_F | 1zhq_A | 2fi9_A | 2p51_A | 2wy7_Q | 3cqt_A | 3hpc_X | 3nzn_B |
| 1g61_A | 1n13_G | 1u07_A | 1zhv_A | 2fj8_A | 2p54_A | 2wy8_A | 3cry_A | 3hr0_A | 3o07_A |
| 1g6a_A | 1n1j_A | 1u09_A | 1zhx_A | 2fl4_A | 2p57_A | 2wya_A | 3csk_A | 3hra_A | 3o0a_B |
| 1g6c_B | 1n1j_B | 1u11_A | 1zi9_A | 2flh_B | 2p5k_A | 2wyq_A | 3ct1_A | 3hrx_A | 3o0d_C |
| 1g6g_A | 1n3y_A | 1u1w_B | 1zja_A | 2fli_A | 2p5y_A | 2wz9_A | 3ctg_A | 3hs3_A | 3o0g_D |
| 1g6h_A | 1n5w_D | 1u2b_A | 1zjc_A | 2fm6_A | 2p65_A | 2wzm_A | 3ctk_A | 3hsh_A | 3o1g_A |
| 1g6u_B | 1n63_C | 1u2h_A | 1zjj_B | 2fma_A | 2p68_A | 2wzx_A | 3ctp_B | 3hss_B | 3o1k_B |
| 1g8a_A | 1n63_E | 1u2p_A | 1zk4_A | 2fmm_C | 2p6h_B | 2x0k_A | 3ctz_A | 3ht1_A | 3o1n_A |
| 1g8i_B | 1n71_C | 1u2w_B | 1zk7_A | 2fmp_A | 2p6w_A | 2x18_E | 3cu4_A | 3ht5_A | 3o1p_A |
| 1g8k_A | 1n7e_A | 1u3i_A | 1zkc_A | 2fn3_A | 2p6z_A | 2x1b_A | 3cu9_A | 3huh_A | 3o26_A |
| 1g8k_D | 1n7s_A | 1u53_A | 1zke_B | 2fn4_A | 2p74_A | 2x23_B | 3cui_A | 3hup_B | 3o2e_A |
| 1g94_A | 1n7s_B | 1u55_A | 1zkk_B | 2fn9_B | 2p8b_A | 2x2o_A | 3cvb_A | 3hv2_B | 3o3m_C |
| 1g97_A | 1n7s_C | 1u5d_B | 1zkl_A | 2fne_A | 2p9x_D | 2x32_A | 3cwn_B | 3hvi_A | 3o3m_D |
| 1g9g_A | 1n7s_D | 1u5f_A | 1zl0_B | 2fnu_A | 2pa1_A | 2x3h_C | 3cwv_A | 3hvu_C | 3o3u_N |
| 1g9o_A | 1n82_B | 1u60_B | 1zlm_A | 2fo3_A | 2pa6_B | 2x3m_A | 3cx5_E | 3hvv_A | 3o4h_A |
| 1ga6_A | 1n83_A | 1u6e_A | 1zm8_A | 2fp1_A | 2pa7_B | 2x49_A | 3cx5_F | 3hx9_A | 3o4r_B |
| 1gai_A | 1n8f_B | 1u6r_B | 1zn8_A | 2fp7_A | 2pbc_C | 2x4d_A | 3cx5_G | 3hxa_F | 3o4v_B |
| 1gbg_A | 1n9l_A | 1u6t_A | 1zo2_A | 2fq3_A | 2pbd_P | 2x4k_B | 3cx5_I | 3hxs_B | 3o5v_B |
| 1gbs_A | 1na0_A | 1u6z_B | 1zoi_B | 2fqw_A | 2pbi_C | 2x4l_A | 3cx5_O | 3hxw_A | 3o70_A |
| 1gci_A | 1na5_A | 1u84_A | 1zos_C | 2fr2_A | 2pbi_D | 2x5c_B | 3cxk_A | 3hz2_A | 3o79_A |
| 1gcq_C | 1nb9_A | 1u8f_Q | 1zps_B | 2fr5_C | 2pbp_A | 2x5f_B | 3cxz_A | 3hzb_E | 3o7b_A |
| 1gde_B | 1nbc_B | 1u8v_C | 1zpw_X | 2frg_P | 2pc8_A | 2x5h_B | 3cy4_A | 3i0w_A | 3o83_B |
| 1gee_E | 1nbu_D | 1u8y_B | 1zq9_B | 2ft0_A | 2pcj_B | 2x5x_A | 3cyi_A | 3i1a_A | 3o85_B |
| 1geg_G | 1nc5_A | 1u9k_A | 1zr0_D | 2ftx_B | 2pcn_A | 2x5y_A | 3cz1_B | 3i1u_A | 3o8m_A |
| 1ges_B | 1ndd_A | 1ua4_A | 1zr3_B | 2fu0_A | 2pdr_B | 2x6w_A | 3czf_B | 3i24_A | 3o9z_A |
| 1gk6_A | 1ne7_C | 1ua6_L | 1zr6_A | 2fu4_A | 2pfz_A | 2x7b_A | 3czt_X | 3i26_D | 3oa3_B |
| 1gk7_A | 1nep_A | 1uai_A | 1zs4_D | 2fuk_A | 2pg0_B | 2x7k_A | 3czz_B | 3i2z_A | 3oaj_A |
| 1gk9_A | 1nf8_A | 1uas_A | 1zsw_A | 2fvh_A | 2pgo_A | 2x7m_A | 3d03_B | 3i31_A | 3oam_A |
| 1gk9_B | 1nff_B | 1ub3_A | 1zsx_A | 2fvv_A | 2ph3_A | 2x8h_A | 3d0n_A | 3i33_A | 3obu_A |
| 1gl2_A | 1nfv_N | 1uc4_A | 1zt5_A | 2fvy_A | 2phn_A | 2x8r_A | 3d0o_A | 3i35_A | 3ocu_A |
| 1gl2_B | 1ng6_A | 1uc4_G | 1zu3_A | 2fwh_A | 2pi6_A | 2x8s_A | 3d1b_C | 3i36_A | 3od9_A |
| 1gl2_C | 1nh2_A | 1uca_A | 1zuo_A | 2fyg_A | 2pie_A | 2x8x_X | 3d1g_A | 3i3f_B | 3odg_A |
| 1gl2_D | 1nh2_B | 1ucr_B | 1zuu_A | 2fyx_A | 2piy_B | 2x96_A | 3d1k_A | 3i3g_A | 3ofk_C |
| 1gmu_C | 1nhc_E | 1ucs_A | 1zuy_A | 2fzp_A | 2pjz_A | 2xb4_A | 3d2q_A | 3i45_A | 3og9_B |
| 1gmy_A | 1nhk_L | 1udc_A | 1zv1_A | 2fzv_B | 2pk3_A | 2xbk_A | 3d2w_A | 3i47_A | 3ogn_B |
| 1gn0_A | 1nki_B | 1ueb_A | 1zwh_A | 2fzw_B | 2pk8_A | 2xbl_A | 3d30_A | 3i48_B | 3ogr_A |
| 1gnl_A | 1nkp_D | 1uek_A | 1zwz_A | 2g2n_C | 2pkf_A | 2xbp_A | 3d32_A | 3i4o_B | 3oid_C |
| 1gnt_A | 1nln_A | 1uf5_A | 1zx6_A | 2g2s_A | 2pko_A | 2xc2_A | 3d34_A | 3i4s_A | 3oig_A |
| 1gny_A | 1nls_A | 1ufb_C | 1zxt_B | 2g30_A | 2pkt_A | 2xcb_A | 3d3b_A | 3i4z_B | 3oiu_A |
| 1go3_N | 1nnf_A | 1ufi_B | 1zxx_A | 2g45_D | 2plt_A | 2xce_F | 3d3z_A | 3i57_B | 3oj7_A |
| 1goi_B | 1nnh_A | 1ufy_A | 1zz0_A | 2g5x_A | 2pmk_A | 2xcj_A | 3d43_B | 3i5c_B | 3ojs_A |
| 1gp6_A | 1nns_A | 1ug6_A | 1zzg_B | 2g64_A | 2pmr_A | 2xcz_A | 3d47_A | 3i5r_A | 3ol0_A |
| 1gpe_A | 1nnw_B | 1ugi_E | 1zzk_A | 2g6f_X | 2pn6_A | 2xda_A | 3d4i_A | 3i5x_A | 3ol3_A |
| 1gpi_A | 1noa_A | 1ugp_B | 1zzo_A | 2g76_A | 2pn8_B | 2xde_A | 3d4u_B | 3i6c_A | 3omc_B |
| 1gpu_A | 1nof_A | 1ugx_A | 1zzw_A | 2g7o_A | 2pnd_A | 2xdg_A | 3d6r_A | 3i6t_B | 3omt_A |
| 1gq1_A | 1nog_A | 1uha_A | 256b_B | 2g84_A | 2pnx_A | 2xdj_F | 3d79_A | 3i7u_B | 3onr_I |
| 1gq8_A | 1nox_A | 1uhe_A | 2a07_K | 2g8o_B | 2pny_A | 2xdp_A | 3d7a_B | 3i8s_C | 3oo8_A |
| 1gql_B | 1npy_B | 1uhk_B | 2a0n_A | 2g9f_A | 2poi_A | 2xdw_A | 3d8t_A | 3i94_A | 3ooi_A |
| 1gqv_A | 1nq6_A | 1ui0_A | 2a14_A | 2ga4_D | 2pok_B | 2xe4_A | 3d95_B | 3i96_A | 3op8_A |
| 1gtf_I | 1nq7_A | 1uiw_C | 2a15_A | 2gag_A | 2por_A | 2xed_A | 3d9t_A | 3i98_E | 3opk_A |
| 1gtv_A | 1nqu_B | 1uj0_A | 2a26_C | 2gag_B | 2pos_D | 2xet_B | 3d9x_C | 3i9q_A | 3oq2_A |
| 1gtz_D | 1nr0_A | 1uj2_A | 2a28_A | 2gag_C | 2ppp_A | 2xeu_A | 3d9y_A | 3ia2_F | 3oqy_B |
| 1gu2_A | 1nr4_G | 1uj8_A | 2a2n_C | 2gas_A | 2ppt_B | 2xev_A | 3da0_C | 3ia4_D | 3orh_C |
| 1gu7_B | 1nsc_B | 1uk7_A | 2a2r_B | 2gb4_B | 2pqm_B | 2xf2_A | 3dac_A | 3ian_A | 3orv_B |
| 1gud_A | 1nth_A | 1ukf_A | 2a40_E | 2gbt_A | 2pqr_B | 2xfd_A | 3dai_A | 3iar_A | 3orv_D |
| 1gug_D | 1ntv_A | 1ukm_A | 2a4v_A | 2gbw_E | 2pqr_D | 2xfv_A | 3dan_A | 3iav_A | 3ose_A |
| 1gui_A | 1nty_A | 1ukm_B | 2a4x_A | 2gc4_L | 2pqx_A | 2xh2_C | 3daq_A | 3ib7_A | 3osm_A |
| 1gv5_A | 1nvm_C | 1uku_A | 2a53_C | 2gdq_B | 2pr5_A | 2xhi_A | 3dau_A | 3ich_A | 3oti_B |
| 1gvj_A | 1nvm_F | 1ukz_A | 2a5d_A | 2gdz_A | 2psd_A | 2xhn_A | 3dc5_C | 3id1_A | 3ouf_B |
| 1gvn_D | 1nw2_H | 1ulk_A | 2a61_C | 2gec_A | 2psp_B | 2xi8_A | 3dcn_A | 3id7_A | 3ovp_B |
| 1gwe_A | 1nwa_A | 1ulr_A | 2a6s_B | 2gey_D | 2pst_X | 2xij_A | 3del_B | 3ida_A | 3oyy_B |
| 1gwi_B | 1nwp_A | 1umd_C | 2a6x_A | 2gf3_B | 2pth_A | 2xio_A | 3deo_A | 3idw_A | 3p0t_A |
| 1gwu_A | 1nww_A | 1umd_D | 2a70_A | 2gf9_A | 2ptt_B | 2xkg_A | 3dfg_A | 3ie4_A | 3p1f_A |
| 1gxn_A | 1nwz_A | 1umk_A | 2a7l_B | 2gg6_A | 2ptz_A | 2xkr_A | 3dgb_A | 3ie5_A | 3p1g_A |
| 1gxu_A | 1nxc_A | 1umz_A | 2a8n_A | 2gh0_B | 2pu3_A | 2xla_A | 3dgp_A | 3iei_C | 3p2n_A |
| 1gxy_A | 1nxj_A | 1uow_A | 2a9i_A | 2gh0_D | 2pu9_A | 2xlk_A | 3dgp_B | 3iev_A | 3p2t_A |
| 1gy6_A | 1nxm_A | 1uoy_A | 2a9s_A | 2gh9_A | 2pu9_B | 2xm5_A | 3dgt_A | 3iez_B | 3p3c_A |
| 1gy7_C | 1nyk_B | 1uqx_A | 2aa1_B | 2gha_A | 2pv2_A | 2xmx_A | 3dha_A | 3ife_A | 3p3e_A |
| 1gyh_C | 1nyt_C | 1uqz_A | 2aal_C | 2gia_B | 2pvb_A | 2xn4_A | 3dhi_B | 3ig9_C | 3p3g_A |
| 1gyo_A | 1nza_A | 1urn_C | 2aan_A | 2gib_B | 2pve_B | 2xn6_A | 3dhi_C | 3igz_B | 3p3o_A |
| 1gyv_A | 1o04_E | 1urr_A | 2ab0_A | 2giy_A | 2pvq_A | 2xov_A | 3dho_C | 3ihw_A | 3p48_A |
| 1gyy_B | 1o0e_B | 1urs_A | 2abk_A | 2gj3_A | 2pwy_A | 2xpp_A | 3die_A | 3ihz_B | 3p4t_A |
| 1gzc_A | 1o1z_A | 1us5_A | 2abw_B | 2gjd_C | 2pxx_A | 2xqu_A | 3dj9_A | 3ii7_A | 3p5h_A |
| 1gzw_B | 1o26_C | 1use_A | 2acf_D | 2gke_A | 2py4_A | 2xs4_A | 3djh_C | 3iij_A | 3p73_A |
| 1h03_P | 1o4k_A | 1usf_B | 2ad6_A | 2gkm_B | 2pyw_A | 2xsu_A | 3djl_A | 3iiu_M | 3p73_B |
| 1h0h_B | 1o4s_A | 1usg_A | 2ad6_D | 2gkr_I | 2pz0_B | 2xt2_A | 3djo_A | 3ij3_A | 3p7y_A |
| 1h16_A | 1o4t_A | 1uso_A | 2ae2_A | 2gl5_A | 2pze_B | 2xts_A | 3dk9_A | 3ijl_A | 3p97_C |
| 1h1n_A | 1o4v_A | 1usq_B | 2aen_E | 2gmy_E | 2pzh_B | 2xts_B | 3dkc_A | 3ik7_D | 3p9c_A |
| 1h1y_A | 1o4y_A | 1uti_A | 2aex_A | 2gn4_B | 2q0l_A | 2xtt_B | 3dkm_A | 3ilo_A | 3p9p_A |
| 1h2b_B | 1o5u_A | 1uu4_A | 2ag4_B | 2gnc_A | 2q20_B | 2xu3_A | 3dkr_A | 3ils_A | 3p9x_A |
| 1h2c_A | 1o5x_A | 1uuq_A | 2ag5_B | 2gok_A | 2q28_A | 2xu8_B | 3dl0_A | 3ilw_A | 3pb6_X |
| 1h2e_A | 1o7e_B | 1uuy_A | 2agd_B | 2gom_A | 2q2a_D | 2xvm_B | 3dlm_A | 3im1_A | 3pbf_A |
| 1h2s_A | 1o7i_A | 1uv4_A | 2ahf_A | 2gou_A | 2q2h_A | 2xvs_A | 3dm8_A | 3im9_A | 3pc3_A |
| 1h2s_B | 1o7j_C | 1uvq_A | 2ahn_A | 2gpe_B | 2q35_A | 2xvx_A | 3dme_B | 3imh_A | 3pcv_A |
| 1h4a_X | 1o7q_B | 1uw4_C | 2aib_A | 2gqt_A | 2q5c_A | 2xws_A | 3dmg_A | 3inz_B | 3pd2_B |
| 1h4g_A | 1o7z_B | 1uw4_D | 2akz_B | 2gqw_A | 2q62_G | 2xwt_C | 3dmi_A | 3iof_A | 3pd7_A |
| 1h4p_A | 1o82_A | 1uwc_A | 2anv_A | 2grc_A | 2q73_C | 2xxj_D | 3dmo_A | 3ioh_A | 3pdn_A |
| 1h4r_A | 1o8s_A | 1uwf_A | 2any_A | 2grr_B | 2q86_B | 2xxl_B | 3dnf_B | 3ioq_A | 3pel_B |
| 1h5b_B | 1o8x_A | 1uwk_B | 2ap1_A | 2gsd_A | 2q87_A | 2xy2_A | 3dpg_B | 3iox_A | 3pew_A |
| 1h5q_L | 1o91_C | 1uwz_A | 2apg_A | 2gso_B | 2q88_A | 2xz2_A | 3dqg_A | 3ip4_A | 3pf2_A |
| 1h5v_A | 1o98_A | 1uxx_X | 2aqm_A | 2gte_A | 2q8n_C | 2xzi_A | 3dr0_C | 3ip8_A | 3pfg_A |
| 1h64_Q | 1o9i_D | 1uxy_A | 2aqp_A | 2gtr_A | 2q8r_G | 2y2z_A | 3dr4_B | 3ipc_A | 3pfs_A |
| 1h6f_A | 1o9r_E | 1uy1_A | 2ar1_A | 2gu3_A | 2q9u_A | 2y39_A | 3dra_A | 3ipf_A | 3pg6_C |
| 1h6l_A | 1oa2_C | 1uyx_A | 2arc_B | 2gud_B | 2qa9_E | 2y3q_B | 3drf_A | 3ipw_A | 3pgx_A |
| 1h6u_A | 1oa8_A | 1uz3_A | 2asd_A | 2gui_A | 2qac_A | 2y3v_D | 3drw_B | 3iq3_A | 3phs_A |
| 1h6w_A | 1oaa_A | 1v05_A | 2asu_B | 2guv_C | 2qap_A | 2y3z_A | 3drz_B | 3iql_A | 3phx_B |
| 1h72_C | 1oai_A | 1v08_B | 2at8_X | 2guy_A | 2qb7_A | 2y5p_C | 3ds4_B | 3irp_X | 3pjp_B |
| 1h75_A | 1oal_A | 1v0z_B | 2atb_A | 2gw4_D | 2qc5_A | 2y7b_A | 3dsk_A | 3irs_A | 3pk0_A |
| 1h7e_B | 1oao_A | 1v2z_A | 2atv_A | 2gwm_A | 2qd6_A | 2y88_A | 3dso_A | 3irv_A | 3pkv_A |
| 1h8p_B | 1oaq_H | 1v30_A | 2au7_A | 2gxg_A | 2qdx_A | 2y8m_A | 3dt9_A | 3is3_A | 3plf_D |
| 1h8u_A | 1oaq_L | 1v33_A | 2avd_B | 2gyq_B | 2qed_A | 2yay_A | 3dtb_A | 3isa_B | 3plw_A |
| 1h97_B | 1obo_A | 1v37_A | 2avk_A | 2gz1_B | 2qee_F | 2ygs_A | 3dvw_A | 3iso_A | 3plx_B |
| 1h98_A | 1oc2_B | 1v4p_C | 2axq_A | 2gz4_A | 2qev_A | 2yqu_B | 3dwg_A | 3isq_A | 3pmc_B |
| 1h9m_A | 1oc8_A | 1v4x_B | 2axw_B | 2gze_A | 2qf4_B | 2yrr_B | 3dwg_C | 3it4_B | 3pmd_A |
| 1h9s_B | 1ock_A | 1v54_A | 2ayd_A | 2gze_B | 2qfa_A | 2ysk_A | 3dwv_B | 3it4_C | 3pms_A |
| 1hbn_C | 1ocy_A | 1v54_J | 2b0a_A | 2gzg_B | 2qfa_B | 2yva_B | 3dxt_A | 3iu5_A | 3pmt_A |
| 1hbn_E | 1odm_A | 1v54_V | 2b0t_A | 2h17_A | 2qfa_C | 2yve_A | 3dy0_A | 3iu7_A | 3pna_A |
| 1hc9_B | 1odt_H | 1v55_D | 2b1k_A | 2h1c_A | 2qfe_A | 2yvi_A | 3dzw_A | 3iux_A | 3po0_A |
| 1hd2_A | 1oe2_A | 1v55_L | 2b3f_D | 2h1v_A | 2qg1_A | 2yvo_A | 3e05_B | 3iwt_A | 3po8_A |
| 1hdo_A | 1off_A | 1v58_B | 2b3h_A | 2h2b_A | 2qgy_B | 2yvt_A | 3e0i_A | 3ix3_B | 3pqa_B |
| 1hfe_L | 1ofl_A | 1v5d_A | 2b49_A | 2h2r_B | 2qhl_B | 2yw2_A | 3e13_X | 3ixq_D | 3pr9_A |
| 1hfe_T | 1ofs_C | 1v5f_A | 2b4z_A | 2h2z_A | 2qho_B | 2yw3_A | 3e17_B | 3jpz_B | 3prp_A |
| 1hfo_E | 1ofs_D | 1v5i_B | 2b5a_A | 2h3h_A | 2qhs_A | 2ywd_A | 3e2d_A | 3jqj_C | 3psm_A |
| 1hfs_A | 1ofw_A | 1v6s_A | 2b5h_A | 2h3l_A | 2qia_A | 2ywj_A | 3e3u_A | 3jql_A | 3pua_A |
| 1hh8_A | 1ofz_A | 1v70_A | 2b5w_A | 2h54_B | 2qif_A | 2ywk_A | 3e4g_A | 3jqu_A | 3pvi_B |
| 1hj8_A | 1ogd_D | 1v7p_C | 2b6n_A | 2h62_A | 2qih_B | 2yxm_A | 3e4w_B | 3jqy_C | 3pxl_A |
| 1hjs_B | 1ogm_X | 1v7r_A | 2b7r_A | 2h64_B | 2qim_A | 2yxn_A | 3e55_A | 3jr0_B | 3q0h_A |
| 1hl7_B | 1oh0_A | 1v7z_F | 2b82_A | 2h6f_A | 2qkh_A | 2yxo_B | 3e6j_A | 3js4_B | 3q12_C |
| 1hle_A | 1oh4_A | 1v8c_C | 2b9w_A | 2h6f_B | 2qkp_C | 2yxw_A | 3e6s_F | 3js5_A | 3q1x_A |
| 1hlq_C | 1oh9_A | 1v8f_A | 2ba2_C | 2h6n_A | 2qlt_A | 2yxz_A | 3e6z_X | 3js8_A | 3q20_B |
| 1hm6_B | 1ohp_B | 1v8h_A | 2bay_E | 2h6u_G | 2qmc_A | 2yyv_B | 3e7d_A | 3jsl_B | 3q23_B |
| 1hml_A | 1oi6_B | 1v93_A | 2bba_A | 2h88_A | 2qmc_D | 2yyy_A | 3e7h_A | 3jsy_B | 3q2e_A |
| 1hmt_A | 1ojk_A | 1v96_B | 2bbe_A | 2h88_B | 2qmm_A | 2yz1_B | 3e7r_L | 3jte_A | 3q3u_A |
| 1hnj_A | 1ojq_A | 1v98_A | 2bcg_G | 2h88_D | 2qmq_A | 2yzc_D | 3e8m_B | 3jtm_A | 3q49_B |
| 1hp1_A | 1ojx_C | 1v9f_A | 2bcg_Y | 2h88_P | 2qn0_A | 2yzh_C | 3e8t_A | 3jtz_A | 3q4t_A |
| 1hpg_A | 1ok0_A | 1vbi_A | 2bcm_B | 2h8e_A | 2qnw_A | 2yzt_A | 3e96_B | 3ju2_A | 3q4u_A |
| 1hq0_A | 1okt_A | 1vbu_A | 2bcr_A | 2h8g_B | 2qo4_A | 2z0a_B | 3e9t_B | 3juu_A | 3q5y_A |
| 1ht6_A | 1on3_A | 1vbw_A | 2bd0_D | 2h9b_A | 2qpn_B | 2z0j_E | 3ea3_B | 3jva_F | 3q62_B |
| 1ht9_B | 1ong_A | 1vc4_B | 2bek_D | 2h9h_A | 2qpw_A | 2z0m_A | 3ea6_A | 3jxo_A | 3q6d_B |
| 1hw1_B | 1oni_C | 1vcd_A | 2bem_A | 2ha8_B | 2qq4_B | 2z0t_C | 3eaz_A | 3jxs_A | 3q6l_A |
| 1hx0_A | 1onj_A | 1vcl_A | 2beq_D | 2hax_B | 2qqi_A | 2z0x_A | 3ebh_A | 3jxy_A | 3q8g_A |
| 1hx1_B | 1ooe_A | 1vd1_A | 2bez_C | 2haz_A | 2qrl_A | 2z1a_A | 3ec0_B | 3jyo_A | 3q93_B |
| 1hx6_C | 1oot_A | 1vd6_A | 2bf6_A | 2hba_A | 2qrw_I | 2z1c_B | 3edf_A | 3jzy_A | 3qan_C |
| 1hxh_D | 1oqj_A | 1ve1_A | 2bfw_A | 2hbv_A | 2qsa_A | 2z2f_A | 3edv_A | 3k01_A | 3qat_B |
| 1hxr_B | 1oqv_C | 1vef_B | 2bgs_A | 2hc8_A | 2qsk_A | 2z38_A | 3ee4_A | 3k1h_A | 3qby_A |
| 1hz4_A | 1orn_A | 1vf1_A | 2bh8_B | 2hc9_A | 2qsq_B | 2z3g_D | 3eeh_A | 3k26_A | 3qc7_A |
| 1hz6_C | 1orr_A | 1vfr_B | 2bii_B | 2hd9_A | 2qt7_B | 2z3v_A | 3ees_A | 3k2w_E | 3qds_B |
| 1hzj_A | 1os6_A | 1vfy_A | 2bjd_B | 2hda_A | 2qub_I | 2z4u_A | 3ef4_A | 3k31_A | 3qe1_A |
| 1hzo_A | 1osy_B | 1vg8_C | 2bjf_A | 2he0_A | 2qud_A | 2z66_B | 3ef6_A | 3k3c_D | 3qgz_A |
| 1hzt_A | 1oth_A | 1vh5_A | 2bji_A | 2he2_A | 2qul_C | 2z6n_A | 3efy_A | 3k3k_A | 3qh4_A |
| 1i0l_A | 1ou8_B | 1vht_B | 2bjq_A | 2he4_A | 2quo_A | 2z6o_A | 3eg4_A | 3k3v_A | 3qhz_M |
| 1i0r_A | 1ouw_C | 1vhw_A | 2bk9_A | 2hek_B | 2quy_H | 2z6r_A | 3egg_D | 3k62_A | 3qk8_C |
| 1i0v_A | 1ov3_A | 1vi6_B | 2bka_A | 2heu_B | 2qv5_A | 2z6w_A | 3ego_B | 3k6f_A | 3qki_B |
| 1i1n_A | 1ow3_A | 1vim_A | 2bkf_A | 2hew_F | 2qvb_A | 2z72_A | 3egw_C | 3k6i_A | 3qmd_A |
| 1i24_A | 1owf_A | 1vj2_A | 2bkl_B | 2hf9_A | 2qvo_A | 2z79_B | 3ehg_A | 3k6v_A | 3qp4_A |
| 1i27_A | 1ox0_A | 1vjk_A | 2bkm_B | 2hfn_H | 2qvu_B | 2z7f_E | 3ehw_B | 3k6y_A | 3qqi_B |
| 1i2t_A | 1oxj_A | 1vjw_A | 2bko_A | 2hgx_B | 2qwc_A | 2z7f_I | 3ei9_B | 3k7f_B | 3qry_A |
| 1i4u_A | 1oxs_C | 1vk5_A | 2bkr_A | 2hhv_A | 2qwl_A | 2z84_A | 3eif_A | 3k7i_B | 3qu1_B |
| 1i6m_A | 1oyg_A | 1vkc_A | 2bkx_A | 2hin_B | 2qwo_B | 2z8f_A | 3ej9_B | 3k7p_B | 3qug_A |
| 1i77_A | 1oz9_A | 1vke_F | 2bky_B | 2hjv_A | 2qx8_B | 2z8l_A | 3ej9_C | 3k89_A | 3qxc_A |
| 1i7h_A | 1ozn_A | 1vki_A | 2bky_Y | 2hke_B | 2qxi_A | 2z8q_A | 3eja_A | 3k8d_A | 3qy1_B |
| 1i7k_A | 1ozw_B | 1vkk_A | 2bl0_A | 2hl7_A | 2qy1_B | 2z8u_B | 3ejf_A | 3k8u_A | 3qyj_A |
| 1i8a_A | 1p0f_B | 1vl1_A | 2bl0_B | 2hlc_A | 2qy9_A | 2z8x_A | 3ejg_A | 3k8w_A | 3qzb_A |
| 1i8f_F | 1p1j_A | 1vl7_A | 2bl8_B | 2hls_A | 2qzt_B | 2z9v_B | 3eju_A | 3k9o_A | 3r0p_B |
| 1i8k_B | 1p1m_A | 1vlc_A | 2blf_A | 2hlv_A | 2r0b_A | 2za0_A | 3eki_A | 3k9w_A | 3r1i_A |
| 1i8o_A | 1p1x_B | 1vlj_A | 2blf_B | 2hmq_D | 2r0h_C | 2zbo_A | 3elw_A | 3kbf_A | 3r1w_C |
| 1i9c_A | 1p28_B | 1vm9_A | 2bme_B | 2hor_A | 2r16_A | 2zbt_B | 3elx_A | 3kcc_A | 3r3r_A |
| 1iap_A | 1p3c_A | 1vmb_A | 2bmo_A | 2hos_B | 2r1j_R | 2zc8_A | 3em1_A | 3kcg_H | 3r3s_C |
| 1iby_B | 1p6o_B | 1vmf_C | 2bmo_B | 2hq6_A | 2r2y_A | 2zd1_A | 3emi_A | 3kci_A | 3r6f_A |
| 1idp_A | 1p71_A | 1vmj_A | 2bnm_B | 2hqh_C | 2r31_A | 2zdh_A | 3emw_A | 3kcp_A | 3sil_A |
| 1ig3_A | 1p99_A | 1vp2_A | 2bo1_A | 2hqs_H | 2r37_A | 2zdo_B | 3enb_A | 3kda_A | 4ubp_A |
| 1igq_A | 1pa2_A | 1vp6_C | 2bo4_F | 2hqy_A | 2r5o_B | 2zdr_A | 3enk_B | 3ke4_A | 4ubp_B |
| 1ihj_B | 1pam_B | 1vph_E | 2bo9_C | 2hra_A | 2r6j_B | 2zex_A | 3enu_A | 3kef_B | 4vub_A |
| 1iib_B | 1pcf_C | 1vps_A | 2bo9_D | 2hrv_B | 2r75_1 | 2zez_B | 3eoi_A | 3keo_B | 5pal_A |
| 1ijb_A | 1pdo_A | 1vq3_B | 2boo_A | 2hsa_A | 2r8e_E | 2zfc_B | 3epr_A | 3kfa_A | 6cel_A |
| 1ijt_A | 1pe9_B | 1vqe_A | 2bpd_B | 2ht9_B | 2r8o_A | 2zfd_A | 3eqn_B | 3kff_A | 6rxn_A |
| 1ijx_C | 1pfb_A | 1vsr_A | 2bpq_A | 2hta_A | 2r8q_A | 2zfz_D | 3er6_A | 3kg0_C | 7fd1_A |
| 1ijy_B | 1pgv_A | 1vyf_A | 2bqx_A | 2hu9_A | 2r99_A | 2zgq_A | 3era_B | 3kgr_A | 7rsa_A |
| 1ikt_A | 1pj5_A | 1vyo_A | 2br9_A | 2hur_B | 2r9f_A | 2zhj_A | 3erj_A | 3kgz_B | 8abp_A |
| 1io0_A | 1pk3_B | 1vzi_B | 2bsj_A | 2hv8_A | 2ra3_B | 2zhn_A | 3erx_B | 3kh7_A | |
| 1iom_A | 1pkh_A | 1vzy_B | 2bt6_A | 2hv8_E | 2ra4_A | 2zhz_C | 3esg_B | 3kij_C | |
| 1ioo_B | 1pl3_A | 1w0d_A | 2bt9_A | 2hvm_A | 2ra6_B | 2zib_A | 3esl_B | 3kki_A | |
| 1iq6_B | 1pl8_D | 1w0n_A | 2buu_A | 2hvw_C | 2rbk_A | 2zjd_C | 3eu9_C | 3kkq_A | |